\documentclass[preprint]{aastex}
\usepackage{natbib}
\shorttitle{ Pitch Angle of Self-Gravity Wakes}
\shortauthors{Michikoshi, Fujii, Kokubo, \& Salo}
\keywords{planets: rings, methods: $N$-body simulations}

\begin{document}

\title{
Dynamics of Self-Gravity Wakes in Dense Planetary Rings I. Pitch Angle
}
\author{
Shugo Michikoshi\altaffilmark{1}, Akihiko Fujii \altaffilmark{2}, 
Eiichiro Kokubo\altaffilmark{1}, and Heikki Salo\altaffilmark{3} 
} 
\altaffiltext{1}{
Division of Theoretical Astronomy, National Astronomical Observatory of
Japan, Osawa, Mitaka, Tokyo 181-8588, Japan 
}
\altaffiltext{2}{
University of Calgary, Calgary, Alberta, Canada T2N 1N4
}
\altaffiltext{3}{
Department of Physics, Astronomy Division, University of Oulu, PO Box 3000, FI-90014, Finland
}

\email{
shugo.michikoshi@nao.ac.jp, akihiko.fujii2@ucalgary.ca, kokubo@th.nao.ac.jp, and heikki.salo@oulu.fi
}

\begin{abstract}

We investigate the dynamics of self-gravity wakes in dense planetary rings.   
In particular, we examine how the pitch angle of self-gravity wakes depend on ring parameters using $N$-body simulations.
We calculate the pitch angles using the two-dimensional autocorrelation function of the ring surface density.
We obtain the pitch angles for the inner and outer parts of the autocorrelation function separately.
We confirm that the pitch angles are 15 to 30 degrees for reasonable ring parameters, which are consistent with previous studies.
We find that the inner pitch angle increases with the Saturnicentric distance, while it barely depends on the optical depth and the restitution coefficient of ring particles. 
The increase of the inner pitch angle with the Saturnicentric distance is consistent with the observations of the A ring. 
The outer pitch angle  does not have the clear dependence on any ring parameters and is about $10 \mbox{--} 15$ degrees.
This value is consistent with the pitch angle of spiral arms in collisionless systems.

\end{abstract}

\section{Introduction}
The main rings of Saturn are composed of numerous icy particles and exhibit many different patterns, including 
non-axisymmetric structures on sub-kilometer scales.
Gravitational forces between particles form gravitationally bound clumps, while differential rotation tears them apart.
Due to these competing processes, spatial structures called gravitational or self-gravity wakes appear
 \citep{Colombo1976, Salo1992a}.
 \cite{Salo1992a} performed the first investigations of self-gravity wakes using $N$-body simulations.
The typical radial scale of the wakes is comparable to the critical
 wavelength of gravitational instability \citep{Toomre1964},
 which is around several tens to hundred meters.
Their azimuthal scale is much longer than their radial scale.

Even though individual wakes are tiny structures, together they can have detectable influence on the large-scale photometric properties of the rings.
The A and B rings exhibit a remarkable asymmetric brightness variation
 that was first discovered by \cite{Camichel1958}, confirmed later by
 the ground-based observations \citep{Ferrin1975, Reitsema1976}, 
 and by the Voyager image data \citep{Franklin1987, Dones1993}.
\cite{Salo2003} and \cite{Salo2004} carried out the detailed photometric
 modeling of self-gravity wakes with $N$-body simulation.
They found that the self-gravity wakes in $N$-body simulation
 can account for the brightness dependence on the longitude and
 elevation in the Voyager and ground-based observations.
This conclusion was confirmed by the Hubble Space Telescope and Cassini
 observations \citep{French2007, Porco2008}.

 Though there have been many numerical $N$-body simulations of self-gravity wakes
 \citep{Salo1992, Richardson1994, Daisaka1999, Daisaka2001, Robbins2010,
 Yasui2012}, the theoretical framework for dense planetary rings with
 self-gravity wakes has not been explored in detail yet.
It is often argued that wakes in dense rings are dynamically similar to
 the spiral structure in differentially rotating galactic disks
 \citep{Colombo1976, Franklin1987}.  
\cite{Julian1966} studied theoretically the spiral structure in disk
 galaxies.  
They found that when disk surface density is perturbed, non-axisymmetric
 density enhancements appear.
This mechanism is called swing amplification \citep{Toomre1981}.
 Based on the swing amplification, we can estimate the pitch angle of a spiral arm \citep{Julian1966, Toomre1981}.  The pitch angle is the angle between the tangents to a spiral arm and a perfect circle, which measures how tightly the spiral arm is wound. 
The swing amplification model predicts that the pitch angle for Keplerian rotation is about $10 \mbox{--} 15^\circ$, which is roughly consistent with that of the wakes in rings.   
Thus, it has been considered that the swing amplification plays an important role in the self-gravity wake formation in rings. 

According to the swing amplification model, the pitch angle of the
 self-gravity wakes depends on the shear rate \citep{Julian1966,
 Michikoshi2014}.  
 However, the dependence of the pitch angle on other ring parameters is still not well understood. 
 Observationally \cite{Hedman2007} investigated the dependence of the pitch angle on the Saturnicentric distance.
They found that the pitch angle varies systematically with the Saturnicentric distance.
In the A ring, the pitch angle is $18^\circ \mbox{--} 24^\circ$, which is consistent with the previous numerical simulations \citep{Salo2004}.
In the middle A ring, the pitch angle is almost constant $18^\circ$, but in the outer A ring, it increases with the Saturnicentric distances to $24^\circ$.
In addition, \cite{Hedman2007} pointed out that the pitch angle has the local maximum at the location of the strong density waves.
From the Cassini UVIS occultation studies, \cite{Colwell2007} obtained the consistent results that the pitch angle in the A ring is $10 \mbox{--} 50 ^\circ$ and it increases with the Saturnicentric distance.  
They also found that the pitch angle in the B ring decreases with the Saturnicentric distance. 
From the HST observations, \cite{French2007} showed the increase of the pitch angle with the Saturnicentric distance in the A ring, which is consistent with the results in \cite{Colwell2007}.  
The pitch angle dependence on the Saturnicentric distance has a complex structure.
It is not a monotonic function of the Saturnicentric distance.
However, roughly speaking, the pitch angle decreases with the Saturnicentric distance in the B ring, while it increases in the A ring.

\cite{Salo1995} performed $N$-body simulations and found that the wake pitch angle is about $20^\circ$ from the autocorrelation function. 
\cite{Salo2004} found that the pitch angle varies with the autocorrelation distance.
The pitch angle for small autocorrelation distance is about $25 \mbox{--} 30^\circ$, and decreases with the distance towards $15 \mbox{--}20^\circ$.
Using the photometric model they found that the longitude of the minimum
 brightness with respect to ansa increases from $17^\circ$ to $23^\circ$ with the elevation angle of the observer with respect to the ring plane.
 They argued that the observed minimum longitude for high elevation angles corresponds to the pitch angle for small distance, and that for low elevation angles corresponds to the pitch angle for large distance since the brightness for high elevation angles is dominated by the inner dense part of the wakes. 
However, the systematic study of the dependence of the pitch angle on
 ring parameters has not been conducted yet. 
 Therefore, we perform such a survey with high resolution $N$-body resolutions.
We calculate the pitch angle from the autocorrelation function of the
 ring surface density and investigate its dependence on the restitution
 coefficient of ring particles, the dynamical optical depth, and the Hill radius of a particle pair relative to the sum of their physical radii. 

The outline of this paper is as follows. 
We describe the simulation model in section 2 and explain the
calculation method of the pitch angle in section 3. 
In section 4, we present the results of the dependence of the pitch
 angle on ring parameters. 
Sections 5 and 6 are devoted for discussion and a summary,
 respectively.

\section{Simulation Method}

\subsection{Equation of Motion}

We adopt the shearing box method developed for local ring dynamics
\citep[e.g.,][]{Wisdom1988, Michikoshi2011}. 
We introduce the rotating local Cartesian coordinates $(x, y, z)$. 
The $x$-axis points radially outward, the $y$-axis points in the
direction of orbital motion, and the $z$-axis points to the normal to the
 ring plane. 
We assume $m_i \ll M_\mathrm{S}$, $|x_i|,|y_i|,|z_i| \ll a$, and
 $|\mathrm{d} x_i/ \mathrm{d}t|,|\mathrm{d} y_i/\mathrm{d}t|,|\mathrm{d}
 z_i/\mathrm{d}t|\ll a\Omega$, where $M_\mathrm{S}$ is the Saturn mass,
 $m_i$ is the mass of $i$th ring particle, $a$ is the Saturnicentric
 distance, and $\Omega$ is the orbital frequency. 
Then the equation of motion for the $i$th particle is
 \cite[e.g.,][]{Petit1986, Nakazawa1988} 
\begin{eqnarray}
  \frac{\mathrm{d}^2  x_i}{\mathrm{d}  t^2} & = &
   2 \Omega \frac{\mathrm{d}  y_i}{\mathrm{d}  t} + 3  \Omega^2 x_i +
   \sum_{j \ne i} \frac{G m_j}{ r_{ij}^3} (x_j -  x_i), \label{eq:eomx} \\
 \frac{\mathrm{d}^2  y_i}{\mathrm{d} t^2} & = &
  -2 \Omega \frac{\mathrm{d} x_i}{\mathrm{d}  t} + 
  \sum_{j \ne i} \frac{G m_j}{ r_{ij}^3} (y_j -  y_i), \label{eq:eomy} \\
  \frac{\mathrm{d}^2 z_i}{\mathrm{d}  t^2} & = & 
   - \Omega^2 z_i  + 
   \sum_{j \ne i} \frac{G m_j}{ r_{ij}^3} (z_j - z_i), \label{eq:eomz}  
\end{eqnarray}
 where $r_{ij}$ is the distance between $i$th and $j$th particles. 
The first terms in Equations (\ref{eq:eomx}) and (\ref{eq:eomy}) are the
 Coriolis force.  
 The second term in Equation (\ref{eq:eomx}) is the tidal force, the first term in
 Equation (\ref{eq:eomz}) is the vertical force due to Saturn's gravity, and the last terms in
 Equations (\ref{eq:eomx})--(\ref{eq:eomz}) are the mutual gravity
 between ring particles. 

The simulation box has a square shape with length $L$ on the $x$-$y$ plane. 
We use the sliding cell method where when a particle crosses a box
boundary outward, the corresponding particle immediately comes into the
 box from the opposite boundary \citep{Wisdom1988, Toomre1991}. 
The length $L$ should be sufficiently larger than the characteristic
 scale length, the critical wavelength of the gravitational instability
\begin{equation}
  \lambda_\mathrm{cr} = \frac{4 \pi^2 G \Sigma_0}{\Omega^2},
\end{equation}
 where $\Sigma_0$ is the mean surface density \citep{Toomre1964}.
We adopt $L = 10 \lambda_\mathrm{cr}$.

For the collision model we adopt the constant restitution coefficient
 $\epsilon$ that ranges from 0.1 to 0.6.
For $\epsilon > 0.7$, the velocity dispersion of ring particles
 increases monotonically and self-gravity wakes do not appear
 \citep{Goldreich1978}.  
The effect of the velocity-dependent restitution coefficient is
 discussed in Section \ref{seceps}.   
For the velocity-dependent restitution coefficient model, we use the
 standard model of \cite{Bridges1984}: 
\begin{equation}
  \epsilon(v_\mathrm{n}) = \min(1,(v_\mathrm{n}/v_\mathrm{c})^{-0.234}),
\end{equation}
 where $v_\mathrm{n}$ is the normal component of the relative velocity
 in collisions and $v_\mathrm{c}=0.0077 \, \mathrm{cm}\, \mathrm{s}^{-1}$. 
 We adopt the hard-sphere model as the collision model \citep[e.g.,][]{Wisdom1988, Salo1991, Richardson1994}. 
   In the hard-sphere model, when collision between particles is detected, the particle velocities change instantly.
   In this study since we neglect the friction between particles, the relative tangential velocity is conserved in collision 
   and the effects of spin are not considered.

 The equation of motion is integrated with a leapfrog integrator with timestep $\Delta t = (2\pi/\Omega)/200$. 
   We calculate the self-gravity by direct summation using the special-purpose computer, GRAPE-DR \citep{Makino2007}.
   We introduce the cutoff of self-gravity by the subregion method \citep{Daisaka1999}.
   In this method, the cutoff length corresponds to the box size $L$.

\subsection{Ring Model}

Ring particles are assumed to be identical for the sake of simplicity
 except for Section \ref{secsize}.
For the identical particle model, all particles have the same mass
 $m_i=m_\mathrm{p}$.  
This assumption is unrealistic since many observations show a power-law
 size distribution of ring particles \citep{Zebker1985,French2000}.  
The effect of the realistic size distribution is discussed in Section
 \ref{secsize} where we adopt the power-law size distribution 
 \citep{Zebker1985,French2000} 
\begin{equation}
  \frac{\mathrm{d}N}{\mathrm{d}R_\mathrm{p}} \propto R_\mathrm{p}^{-q},
  \label{eq:size}
\end{equation}
 where $R_\mathrm{p}$ is the particle radius and $q$ is the power-law index.
 From the Voyager radio occultation measurements, the power-law index increases with the Saturnicentric distance from $2.7$ to $3.0$ in the A ring \citep{Zebker1985}.
   The stellar occultation observations showed the consistent results that the power-law indexes are $q=2.75$ and $2.9$ in the inner and outer A rings, respectively \citep{French2000}. Considering these results, we choose the intermediate value $q=2.8$. 
We set the lower and upper size limits as 
 $R_\mathrm{min} = 150 /\sqrt{\gamma} \, \mathrm{cm}$ and
 $R_\mathrm{max} = 150 \sqrt{\gamma} \, \mathrm{cm}$ where 
 $\gamma = R_\mathrm{max}/R_\mathrm{min} = 1\mbox{--}10$.

 For a ring with identical particles, the dynamical
 optical depth is given by 
\begin{equation}
  \tau = \frac{N \pi R_\mathrm{p}^2}{L^2} = \frac{3 \Sigma_0}{4 R_\mathrm{p} \rho_\mathrm{p}} = 0.417
  \left(\frac{\Sigma_0}{50 \, \mathrm{g} \, \mathrm{cm}^{-2}} \right)
  \left(\frac{R_\mathrm{p}}{100 \, \mathrm{cm}} \right)^{-1} 
  \left(\frac{\rho_\mathrm{p}}{0.9\, \mathrm{g}\, \mathrm{cm}^{-3}} \right)^{-1},
  \label{eq:tau}
\end{equation}
 where $N$ is the number of particles in the simulation box
 and $\rho_\mathrm{p}$ is the particle density. 
 For a low optical depth ring, the dynamical optical depth is
 equal to the observed optical depth \citep{Salo2003,Robbins2010}, while
 for a high optical depth ring, it is larger than the observed optical depth.
Though the particle density of solid ice is $0.9 \, \mathrm{g}\,
 \mathrm{cm}^{-3}$, $\rho_\mathrm{p}$ is estimated to be $0.45 \,
 \mathrm{g}\, \mathrm{cm}^{-3}$ from the best fit to the observations of azimuthal brightness variation in the A ring \citep{Salo2004, French2007, Porco2008}. 
The surface density is inferred from density waves, which is typically
 $\Sigma_0 \simeq 45 \, \mathrm{g} \, \mathrm{cm}^{-2}$ in the A ring
 \citep[e.g.][]{Tiscareno2007}. 
The surface density, radius, and particle density of ring particles
 collectively affect the structure of self-gravity wakes. 
 We vary the dynamical optical depth from 0.3 to 1.8 in the identical particle models, which corresponds to the surface density of $\Sigma_0 = 36 \mbox{--} 216 \, \mathrm{g}\, \mathrm{cm}^{-2}$ with $ R_\mathrm{p} = 100 \, \mathrm{cm}$ and $\rho_\mathrm{p} = 0.9\, \mathrm{g}\, \mathrm{cm}^{-3}$.

Both the semi-major axis and particle density affect the spatial structure. 
If the distance of the simulation box from the planet, i.e., the
 semi-major axis is large, the tidal force becomes less effective,
 allowing ring particles to gravitationally accumulate.
In the region outside the Roche limit, particles can form
 gravitationally-bound aggregates.  
The particle density affects the strength of the
 self-gravity. 
Larger density makes particles pull each other more strongly, causing more
 distinct wakes.  
These factors are combined into one non-dimensional parameter, 
the Hill radius of a particle pair relative to the sum of their physical radii, defined as 
\begin{equation}
 \tilde r_\mathrm{H} \equiv 
 \frac{R_\mathrm{H}}{2R_\mathrm{p}} = 
 0.82 
 \left(\frac{M_\mathrm{S}}{5.69\times 10^{29} \, \mathrm{g}} \right)^{-1/3}
 \left(\frac{\rho_\mathrm{p}}{0.9\, \mathrm{g}\, \mathrm{cm}^{-3}} \right)^{1/3}
 \left(\frac{a}{10^{5}\, \mathrm{km}} \right),
 \label{eq:rhdif}
\end{equation}
 where $R_\mathrm{H}$ is the Hill radius of the particle 
 $R_\mathrm{H} = (2m_\mathrm{p}/3M_\mathrm{S})^{1/3}a$. 
With the particle density $\rho_\mathrm{p}$ fixed, $\tilde r_\mathrm{H}$
 is proportional to $a$. 
We vary the ratio from 0.68 to 1.06.

We introduce a parameter, initial Toomre $Q$ to determine the initial
 random velocity of particles,  
\begin{equation}
 Q_\mathrm{ini}= \frac{\sigma_x \Omega}{3.36 G \Sigma_0},
 \label{eq:qval}
\end{equation}
 where $\sigma_x$ is the initial radial velocity dispersion
 \citep{Toomre1964}. 
Since the ring relaxes into the stationary state quickly, $Q_\mathrm{ini}$ 
does not affect the final state \citep{Salo2012}.
Thus, we adopt $Q_\mathrm{ini} = 2$.

The ring models are summarized in Table \ref{tbl:model1} for identical
 particles and in Table \ref{tbl:model2} for particles with size
 distributions.
 Note that the dynamical properties of rings with the identical particles are characterized by the three non-dimensional quantities: $\tau$, $\tilde r_\mathrm{H}$ and $\epsilon$ \citep[e.g.,][]{Karjalainen2004}. The dimensional quantities such as particle size are irrelevant to simulation results. 
   However, in order to compare with the actual rings, we add the corresponding dimensional quantities to Table \ref{tbl:model1} assuming 
   $R_\mathrm{p} = 100\, \mathrm{cm}$ and $\rho_\mathrm{p}= 0.45 \mbox{--} 0.90 \, \mathrm{g}\,  \mathrm{cm}^{-3}$.
The surface density $\Sigma_0$ is calculated by Equation
 (\ref{eq:tau}) and the Saturnicentric distance $a$ is calculated by Equation (\ref{eq:rhdif}).

\begin{deluxetable}{cccccccccccc}
  \tabletypesize{\scriptsize}
  \tablecaption{Ring Parameters and Pitch Angles for Identical Particle Models\label{tbl:model1}}
  \tablewidth{0pt}        
  \tablehead{             
	\colhead{Model} & \colhead{$\tau$} & \colhead{$\tilde r_\mathrm{H}$} & \colhead{$\epsilon$} & \colhead{$N$} 
	& \colhead{$\Sigma_0$ }
	& \colhead{$a$ }
	& \colhead{$L$ }
	& \colhead{$\theta_\mathrm{i}$ }
	& \colhead{$\theta_\mathrm{o}$ }
	& \colhead{$x_\mathrm{b}$}
	& \colhead{$|y_\mathrm{b}|$}\\
	\colhead{} & 
	\colhead{} & 
	\colhead{} & \colhead{} & 
	\colhead{} 
	& \colhead{($\mathrm{g\,cm^{-2}}$)}
	& \colhead{($10^{5}\, \mathrm{km}$)}
	& \colhead{($\mathrm{km}$)}
	& \colhead{(deg.)}
	& \colhead{(deg.)}
	& \colhead{($\lambda_\mathrm{cr}$)}
	& \colhead{($\lambda_\mathrm{cr}$)}
  }                       
  \startdata              
  1 & 0.60 & 0.81 & 0.50 & 44156 & $36\mbox{--}72$ & $0.78\mbox{--}0.99$ & 0.48 & $25.8\pm4.1$ & $13.3\pm4.2$ & 0.26 & 0.53 \\
 &  &  &  & &  &  &  &  &  &  &  \\
  2 & 0.70 & 0.81 & 0.10 & 70119 & $42\mbox{--}84$ & $0.78\mbox{--}0.99$ & 0.56 & $27.3\pm6.0$ & $10.9\pm6.2$ & 0.28 & 0.54 \\
  3 & 0.70 & 0.81 & 0.20 & 70119 & $42\mbox{--}84$ & $0.78\mbox{--}0.99$ & 0.56 & $25.7\pm4.3$ & $15.4\pm7.0$ & 0.27 & 0.55 \\
  4 & 0.70 & 0.81 & 0.30 & 70119 & $42\mbox{--}84$ & $0.78\mbox{--}0.99$ & 0.56 & $26.0\pm4.3$ & $13.8\pm5.4$ & 0.31 & 0.63 \\
  5 & 0.70 & 0.81 & 0.40 & 70119 & $42\mbox{--}84$ & $0.78\mbox{--}0.99$ & 0.56 & $25.7\pm3.9$ & $15.6\pm4.7$ & 0.23 & 0.49 \\
  6 & 0.70 & 0.81 & 0.50 & 70119 & $42\mbox{--}84$ & $0.78\mbox{--}0.99$ & 0.56 & $25.2\pm3.1$ & $15.1\pm5.3$ & 0.26 & 0.55 \\
  7 & 0.70 & 0.81 & 0.60 & 70119 & $42\mbox{--}84$ & $0.78\mbox{--}0.99$ & 0.56 & $27.0\pm3.1$ & $14.1\pm3.3$ & 0.14 & 0.28 \\
  8 & 0.70 & 0.81 & $\epsilon(v_n)$ & 70119 & $42\mbox{--}84$ & $0.78\mbox{--}0.99$ & 0.56 & $24.5\pm4.2$ & $13.9\pm4.1$ & 0.27 & 0.60 \\
 &  &  &  & &  &  &  &  &  &  &  \\
  9 & 0.80 & 0.60 & 0.50 & 17290 & $48\mbox{--}96$ & $0.58\mbox{--}0.73$ & 0.26 & $17.9\pm1.5$ & $10.0\pm1.5$ & 0.74 & 2.29 \\
  10 & 1.00 & 0.60 & 0.50 & 33770 & $60\mbox{--}120$ & $0.58\mbox{--}0.73$ & 0.33 & $22.1\pm2.3$ & $13.5\pm1.5$ & 0.04 & 0.10 \\
  11 & 1.20 & 0.60 & 0.50 & 58355 & $72\mbox{--}144$ & $0.58\mbox{--}0.73$ & 0.39 & $9.7\pm2.1$ & $17.5\pm6.1$ & 0.24 & 1.39 \\
  12 & 1.40 & 0.60 & 0.50 & 92666 & $84\mbox{--}168$ & $0.58\mbox{--}0.73$ & 0.46 & $8.2\pm3.7$ & $19.8\pm10.5$ & 0.19 & 1.28 \\
  13 & 1.60 & 0.60 & 0.50 & 138324 & $96\mbox{--}192$ & $0.58\mbox{--}0.73$ & 0.52 & $0.5\pm2.7$ & $-0.3\pm2.3$ & 0.03 & 3.98 \\
  14 & 1.80 & 0.60 & 0.50 & 196950 & $108\mbox{--}216$ & $0.58\mbox{--}0.73$ & 0.59 & $9.8\pm1.4$ & $17.8\pm6.5$ & 0.29 & 1.70 \\
 &  &  &  & &  &  &  &  &  &  &  \\
  15 & 0.30 & 0.81 & 0.50 & 5519 & $18\mbox{--}36$ & $0.78\mbox{--}0.99$ & 0.24 & $30.2\pm3.0$ & $13.5\pm2.6$ & 0.29 & 0.50 \\
  16 & 0.40 & 0.81 & 0.50 & 13083 & $24\mbox{--}48$ & $0.78\mbox{--}0.99$ & 0.32 & $30.3\pm2.5$ & $13.2\pm3.4$ & 0.22 & 0.37 \\
  17 & 0.50 & 0.81 & 0.50 & 25553 & $30\mbox{--}60$ & $0.78\mbox{--}0.99$ & 0.40 & $29.5\pm2.8$ & $12.3\pm4.4$ & 0.21 & 0.37 \\
  18 & 0.70 & 0.81 & 0.50 & 70119 & $42\mbox{--}84$ & $0.78\mbox{--}0.99$ & 0.56 & $25.2\pm3.1$ & $15.1\pm5.3$ & 0.26 & 0.55 \\
  19 & 0.80 & 0.81 & 0.50 & 104667 & $48\mbox{--}96$ & $0.78\mbox{--}0.99$ & 0.64 & $25.9\pm3.9$ & $16.9\pm6.0$ & 0.16 & 0.32 \\
  20 & 0.90 & 0.81 & 0.50 & 149028 & $54\mbox{--}108$ & $0.78\mbox{--}0.99$ & 0.72 & $24.5\pm3.9$ & $12.8\pm5.8$ & 0.35 & 0.76 \\
  21 & 1.00 & 0.81 & 0.50 & 204428 & $60\mbox{--}120$ & $0.78\mbox{--}0.99$ & 0.80 & $24.5\pm4.8$ & $13.4\pm5.7$ & 0.30 & 0.67 \\
  22 & 1.10 & 0.81 & 0.50 & 272094 & $66\mbox{--}132$ & $0.78\mbox{--}0.99$ & 0.88 & $23.4\pm4.1$ & $13.6\pm6.3$ & 0.47 & 1.09 \\
  23 & 1.20 & 0.81 & 0.50 & 353253 & $72\mbox{--}144$ & $0.78\mbox{--}0.99$ & 0.96 & $23.7\pm3.9$ & $11.2\pm6.2$ & 0.54 & 1.23 \\
  24 & 1.30 & 0.81 & 0.50 & 449130 & $78\mbox{--}156$ & $0.78\mbox{--}0.99$ & 1.04 & $24.7\pm4.8$ & $13.6\pm7.3$ & 0.37 & 0.81 \\
 &  &  &  & &  &  &  &  &  &  &  \\
  25 & 0.30 & 0.68 & 0.50 & 1932 & $18\mbox{--}36$ & $0.66\mbox{--}0.83$ & 0.14 & $26.9\pm5.2$ & $8.4\pm5.2$ & 0.80 & 1.58 \\
  26 & 0.30 & 0.73 & 0.50 & 2957 & $18\mbox{--}36$ & $0.71\mbox{--}0.89$ & 0.18 & $26.4\pm3.6$ & $13.5\pm2.6$ & 0.44 & 0.89 \\
  27 & 0.30 & 0.77 & 0.50 & 4073 & $18\mbox{--}36$ & $0.75\mbox{--}0.94$ & 0.21 & $28.1\pm2.9$ & $14.0\pm3.3$ & 0.30 & 0.57 \\
  28 & 0.30 & 0.81 & 0.50 & 5519 & $18\mbox{--}36$ & $0.78\mbox{--}0.99$ & 0.24 & $30.2\pm3.0$ & $13.5\pm2.6$ & 0.29 & 0.50 \\
  29 & 0.30 & 0.85 & 0.50 & 7370 & $18\mbox{--}36$ & $0.82\mbox{--}1.04$ & 0.28 & $32.2\pm2.7$ & $12.5\pm3.2$ & 0.30 & 0.47 \\
  30 & 0.30 & 0.90 & 0.50 & 10386 & $18\mbox{--}36$ & $0.87\mbox{--}1.10$ & 0.33 & $37.1\pm3.8$ & $13.1\pm4.0$ & 0.27 & 0.36 \\
  31 & 0.30 & 0.94 & 0.50 & 13482 & $18\mbox{--}36$ & $0.91\mbox{--}1.15$ & 0.38 & $42.3\pm4.0$ & $13.8\pm4.1$ & 0.24 & 0.26 \\
  32 & 0.30 & 0.98 & 0.50 & 17312 & $18\mbox{--}36$ & $0.95\mbox{--}1.20$ & 0.43 & $47.3\pm5.3$ & $13.3\pm5.0$ & 0.24 & 0.22 \\
  33 & 0.30 & 1.02 & 0.50 & 22008 & $18\mbox{--}36$ & $0.99\mbox{--}1.24$ & 0.48 & $50.4\pm6.0$ & $10.4\pm5.2$ & 0.31 & 0.25 \\
  34 & 0.30 & 1.06 & 0.50 & 27722 & $18\mbox{--}36$ & $1.03\mbox{--}1.29$ & 0.54 & $53.7\pm5.5$ & $7.3\pm4.8$ & 0.41 & 0.30 \\
 &  &  &  & &  &  &  &  &  &  &  \\
  35 & 0.60 & 0.68 & 0.50 & 15457 & $36\mbox{--}72$ & $0.66\mbox{--}0.83$ & 0.28 & $22.1\pm1.3$ & $12.5\pm1.6$ & 0.20 & 0.50 \\
  36 & 0.60 & 0.73 & 0.50 & 23660 & $36\mbox{--}72$ & $0.71\mbox{--}0.89$ & 0.35 & $26.4\pm1.9$ & $13.3\pm2.4$ & 0.11 & 0.22 \\
  37 & 0.60 & 0.77 & 0.50 & 32585 & $36\mbox{--}72$ & $0.75\mbox{--}0.94$ & 0.41 & $25.2\pm2.3$ & $13.4\pm4.0$ & 0.18 & 0.39 \\
  38 & 0.60 & 0.85 & 0.50 & 58965 & $36\mbox{--}72$ & $0.82\mbox{--}1.04$ & 0.56 & $27.9\pm4.3$ & $13.7\pm6.4$ & 0.32 & 0.61 \\
  39 & 0.60 & 0.90 & 0.50 & 83088 & $36\mbox{--}72$ & $0.87\mbox{--}1.10$ & 0.66 & $32.0\pm5.6$ & $9.8\pm6.3$ & 0.40 & 0.64 \\
  40 & 0.60 & 0.94 & 0.50 & 107858 & $36\mbox{--}72$ & $0.91\mbox{--}1.15$ & 0.75 & $35.7\pm5.2$ & $16.7\pm6.0$ & 0.25 & 0.35 \\
  41 & 0.60 & 0.98 & 0.50 & 138497 & $36\mbox{--}72$ & $0.95\mbox{--}1.20$ & 0.85 & $39.0\pm7.6$ & $14.0\pm7.4$ & 0.31 & 0.38 \\
  42 & 0.60 & 1.02 & 0.50 & 176070 & $36\mbox{--}72$ & $0.99\mbox{--}1.24$ & 0.96 & $41.9\pm7.2$ & $14.7\pm6.8$ & 0.34 & 0.37 \\
  43 & 0.60 & 1.06 & 0.50 & 221779 & $36\mbox{--}72$ & $1.03\mbox{--}1.29$ & 1.08 & $46.4\pm8.9$ & $6.7\pm6.1$ & 0.56 & 0.53 \\
 &  &  &  & &  &  &  &  &  &  &  \\
  44 & 0.90 & 0.68 & 0.50 & 52169 & $54\mbox{--}108$ & $0.66\mbox{--}0.83$ & 0.43 & $15.0\pm1.7$ & $13.4\pm1.8$ & 0.21 & 0.78 \\
  45 & 0.90 & 0.73 & 0.50 & 79854 & $54\mbox{--}108$ & $0.71\mbox{--}0.89$ & 0.53 & $21.6\pm2.1$ & $14.8\pm3.1$ & 0.09 & 0.22 \\
  46 & 0.90 & 0.77 & 0.50 & 109977 & $54\mbox{--}108$ & $0.75\mbox{--}0.94$ & 0.62 & $22.6\pm3.2$ & $14.2\pm5.3$ & 0.19 & 0.45 \\
  47 & 0.90 & 0.81 & 0.50 & 149028 & $54\mbox{--}108$ & $0.78\mbox{--}0.99$ & 0.72 & $24.5\pm3.9$ & $12.8\pm5.8$ & 0.35 & 0.76 \\
  48 & 0.90 & 0.85 & 0.50 & 199009 & $54\mbox{--}108$ & $0.82\mbox{--}1.04$ & 0.83 & $26.2\pm6.1$ & $17.7\pm7.1$ & 0.28 & 0.57 \\
  49 & 0.90 & 0.90 & 0.50 & 280423 & $54\mbox{--}108$ & $0.87\mbox{--}1.10$ & 0.99 & $31.2\pm5.2$ & $14.7\pm5.8$ & 0.30 & 0.49 \\
  50 & 0.90 & 0.94 & 0.50 & 364021 & $54\mbox{--}108$ & $0.91\mbox{--}1.15$ & 1.13 & $33.0\pm9.9$ & $9.8\pm7.6$ & 0.49 & 0.76 \\
  51 & 0.90 & 0.98 & 0.50 & 467429 & $54\mbox{--}108$ & $0.95\mbox{--}1.20$ & 1.28 & $39.0\pm8.9$ & $14.2\pm5.7$ & 0.28 & 0.35 \\
  52 & 0.90 & 1.02 & 0.50 & 594238 & $54\mbox{--}108$ & $0.99\mbox{--}1.24$ & 1.44 & $38.8\pm11.6$ & $11.7\pm9.4$ & 0.56 & 0.70 \\
  53 & 0.90 & 1.06 & 0.50 & 748505 & $54\mbox{--}108$ & $1.03\mbox{--}1.29$ & 1.62 & $47.4\pm10.4$ & $14.3\pm5.6$ & 0.29 & 0.27 \\
  \enddata                
\end{deluxetable}

\begin{deluxetable}{cccccccccc}
  \tabletypesize{\scriptsize}
  \tablecaption{Ring Parameters and Pitch Angles for Size Distribution Models \label{tbl:model2}
  }
  \tablewidth{0pt}        
  \tablehead{             
	\colhead{Model} 
	& \colhead{$a$ }
	& \colhead{$\rho_\mathrm{p}$}
	& \colhead{$R_\mathrm{min}\mbox{--}R_\mathrm{max}$}
	& \colhead{$N$}
	& \colhead{$L$}
	& \colhead{$\theta_\mathrm{i}$}
	& \colhead{$\theta_\mathrm{o}$}
	& \colhead{$x_\mathrm{b}$}
	& \colhead{$|y_\mathrm{b}|$}\\
	& \colhead{($10^5\, \mathrm{km}$)}
	& \colhead{($\mathrm{g}\,  \mathrm{cm}^{-3}$)}
	& \colhead{($\mathrm{cm}$)}
	& \colhead{}
	& \colhead{$(\mathrm{km})$}
	& \colhead{(deg.)}
	& \colhead{(deg.)}
	& \colhead{($\lambda_\mathrm{cr}$)}
	& \colhead{($\lambda_\mathrm{cr}$)}
  }                       
  \startdata              
  54 & 1.30 & 0.45 & $150\mbox{--}150$ & 46148 & 0.76 & $30.0\pm5.0$ & $14.8\pm8.3$ & 0.21 & 0.36 \\
  55 & 1.30 & 0.45 & $106\mbox{--}212$ & 47421 & 0.76 & $31.6\pm8.5$ & $15.6\pm6.9$ & 0.18 & 0.30 \\
  56 & 1.30 & 0.45 & $87\mbox{--}260$ & 49392 & 0.76 & $34.3\pm8.0$ & $14.8\pm6.9$ & 0.20 & 0.30 \\
  57 & 1.30 & 0.45 & $75\mbox{--}300$ & 52260 & 0.76 & $34.1\pm8.0$ & $13.7\pm6.4$ & 0.26 & 0.39 \\
  58 & 1.30 & 0.45 & $67\mbox{--}335$ & 54745 & 0.76 & $35.5\pm5.6$ & $13.4\pm5.3$ & 0.27 & 0.38 \\
  59 & 1.30 & 0.45 & $61\mbox{--}367$ & 57128 & 0.76 & $37.5\pm5.6$ & $14.9\pm7.0$ & 0.23 & 0.30 \\
  60 & 1.30 & 0.45 & $57\mbox{--}397$ & 58069 & 0.76 & $41.1\pm8.5$ & $15.8\pm5.9$ & 0.21 & 0.24 \\
  61 & 1.30 & 0.45 & $53\mbox{--}424$ & 60591 & 0.76 & $40.9\pm7.2$ & $14.7\pm5.8$ & 0.24 & 0.28 \\
  62 & 1.30 & 0.45 & $50\mbox{--}450$ & 62207 & 0.76 & $41.1\pm6.9$ & $15.1\pm6.1$ & 0.26 & 0.29 \\
  63 & 1.30 & 0.45 & $47\mbox{--}474$ & 64929 & 0.76 & $41.6\pm5.0$ & $15.1\pm4.7$ & 0.25 & 0.28 \\
   &  &  & &  & &  &  &  &  \\
  64 & 0.85 & 0.45 & $47\mbox{--}474$ & 4998 & 0.21 & $22.0\pm8.9$ & $18.2\pm5.5$ & 0.67 & 1.66 \\
  65 & 0.90 & 0.45 & $47\mbox{--}474$ & 7043 & 0.25 & $28.0\pm5.1$ & $11.0\pm4.2$ & 1.02 & 1.92 \\
  66 & 0.95 & 0.45 & $47\mbox{--}474$ & 9742 & 0.30 & $22.6\pm6.4$ & $18.8\pm3.9$ & 0.36 & 0.87 \\
  67 & 1.00 & 0.45 & $47\mbox{--}474$ & 13254 & 0.35 & $34.2\pm4.7$ & $15.2\pm4.0$ & 0.26 & 0.39 \\
  68 & 1.05 & 0.45 & $47\mbox{--}474$ & 17761 & 0.40 & $30.4\pm5.7$ & $12.8\pm5.2$ & 0.41 & 0.70 \\
  69 & 1.10 & 0.45 & $47\mbox{--}474$ & 23480 & 0.46 & $31.4\pm4.3$ & $10.7\pm5.7$ & 0.50 & 0.82 \\
  70 & 1.15 & 0.45 & $47\mbox{--}474$ & 30657 & 0.53 & $32.8\pm6.5$ & $13.7\pm7.1$ & 0.30 & 0.47 \\
  71 & 1.20 & 0.45 & $47\mbox{--}474$ & 39576 & 0.60 & $34.8\pm5.3$ & $12.5\pm6.6$ & 0.32 & 0.46 \\
  72 & 1.25 & 0.45 & $47\mbox{--}474$ & 50560 & 0.68 & $35.9\pm7.1$ & $14.4\pm5.4$ & 0.29 & 0.40 \\
  73 & 1.30 & 0.45 & $47\mbox{--}474$ & 63974 & 0.76 & $43.0\pm5.7$ & $14.6\pm5.2$ & 0.25 & 0.26 \\
  74 & 1.35 & 0.45 & $47\mbox{--}474$ & 80232 & 0.85 & $41.1\pm8.2$ & $8.8\pm6.4$ & 0.48 & 0.55 \\
   &  &  & &  & &  &  &  &  \\
  75 & 0.85 & 0.90 & $47\mbox{--}474$ & 2499 & 0.21 & $24.8\pm13.9$ & $19.4\pm8.5$ & 0.89 & 1.92 \\
  76 & 0.90 & 0.90 & $47\mbox{--}474$ & 3521 & 0.25 & $36.6\pm11.0$ & $16.0\pm8.5$ & 0.53 & 0.72 \\
  77 & 0.95 & 0.90 & $47\mbox{--}474$ & 4871 & 0.30 & $37.9\pm21.0$ & $14.8\pm10.4$ & 0.34 & 0.44 \\
  78 & 1.00 & 0.90 & $47\mbox{--}474$ & 6627 & 0.35 & $44.3\pm16.2$ & $13.9\pm8.0$ & 0.48 & 0.49 \\
  79 & 1.05 & 0.90 & $47\mbox{--}474$ & 8880 & 0.40 & $44.9\pm13.4$ & $8.6\pm7.9$ & 0.93 & 0.93 \\
  80 & 1.10 & 0.90 & $47\mbox{--}474$ & 11740 & 0.46 & $44.3\pm10.1$ & $15.4\pm9.2$ & 0.42 & 0.43 \\
  81 & 1.15 & 0.90 & $47\mbox{--}474$ & 15328 & 0.53 & $45.6\pm26.1$ & $12.5\pm9.1$ & 0.49 & 0.48 \\
  82 & 1.20 & 0.90 & $47\mbox{--}474$ & 19788 & 0.60 & $54.9\pm7.0$ & $12.6\pm8.2$ & 0.48 & 0.33 \\
  83 & 1.25 & 0.90 & $47\mbox{--}474$ & 25280 & 0.68 & $58.0\pm8.0$ & $12.8\pm6.7$ & 0.51 & 0.32 \\
  84 & 1.30 & 0.90 & $47\mbox{--}474$ & 31987 & 0.76 & $90.4\pm8.8$ & $16.3\pm11.2$ & 0.22 & 0.00 \\
  85 & 1.35 & 0.90 & $47\mbox{--}474$ & 40116 & 0.85 & $90.2\pm2.9$ & $16.2\pm8.7$ & 0.28 & 0.00 \\
  \enddata                
  \tablecomments{The surface density is $\Sigma_0 = 50\, \mathrm{g}\, \mathrm{cm}^{-2}$ and the restitution coefficient is $\epsilon = 0.5$.  }
\end{deluxetable}

\section{Structure of Self-Gravity Wakes}

\subsection{Formation of Self-Gravity Wakes}

\begin{figure}
 \begin{minipage}{0.33\hsize}
 \begin{center} (a) $\tilde t = 0.0$ \end{center}
   \includegraphics[width=\columnwidth]{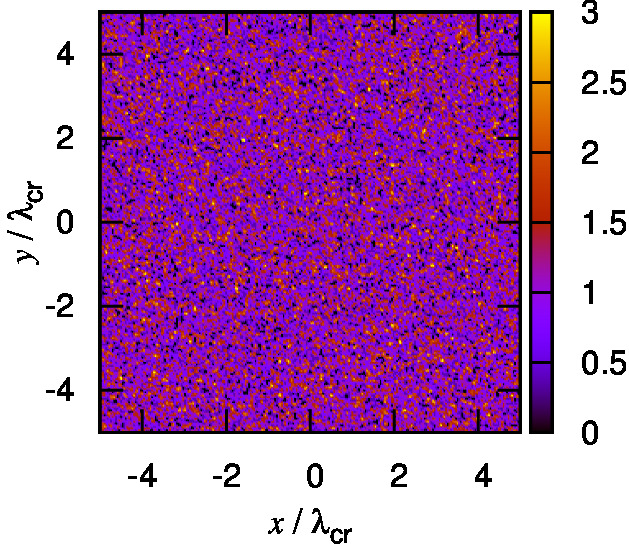}
 \end{minipage}
 \begin{minipage}{0.33\hsize}
   \begin{center} (b) $\tilde t = 1.0$ \end{center}
   \includegraphics[width=\columnwidth]{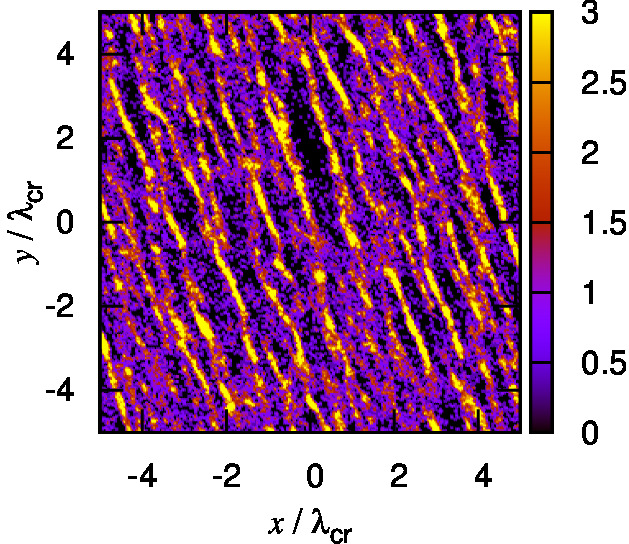}
 \end{minipage}
 \begin{minipage}{0.33\hsize}
 \begin{center} (c) $\tilde t = 5.0$ \end{center}
   \includegraphics[width=\columnwidth]{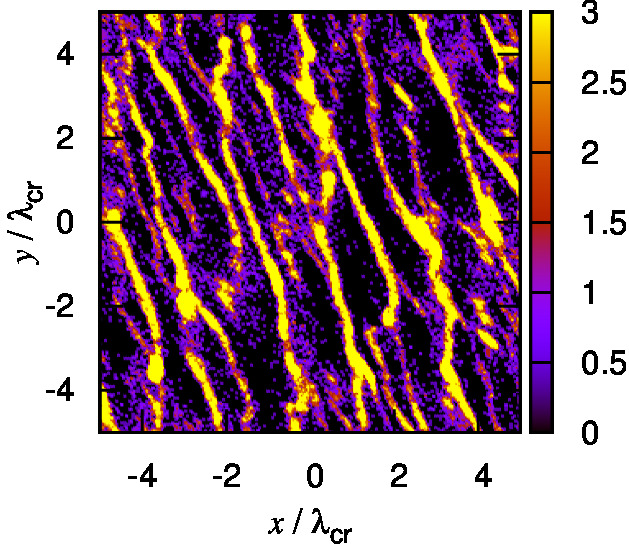}
 \end{minipage}
 \caption{
Snapshots of the surface density normalized by the mean surface density $\Sigma_0$ at $\tilde t=0.0$ (left),
 $\tilde t=1.0$ (middle) and $\tilde t=5.0$ (right) for model 1. 
}
\label{fig:timeevo_of_snapshot}
\end{figure}

Figure \ref{fig:timeevo_of_snapshot} presents an example of the
 self-gravity wake formation for the standard model (model 1). 
The parameters are $\tau=0.6$, $\tilde r_\mathrm{H}=0.81$, and $\epsilon=0.5$.
Initially particles are distributed uniformly in the simulation box 
 (Fig.\ref{fig:timeevo_of_snapshot}a).

\begin{figure}
  \plotone{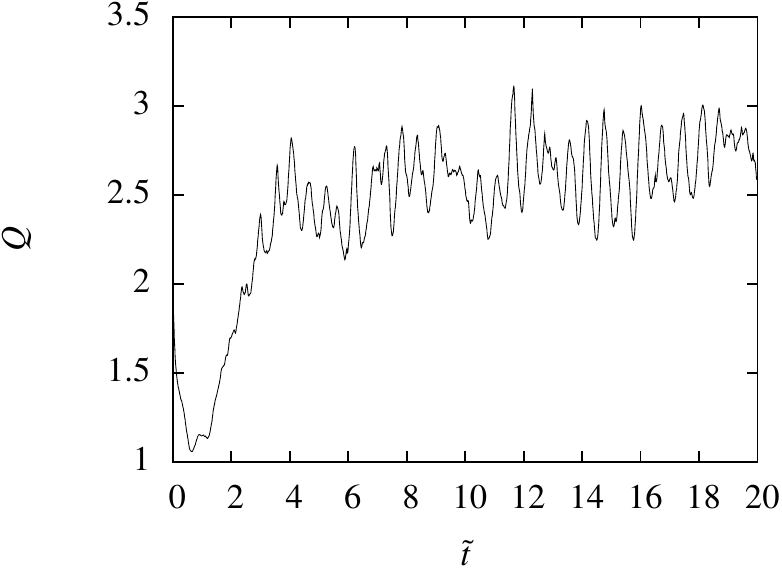}
  \caption{Time evolution of $Q$ for model 1.
}
\label{fig:qvalueevo}
\end{figure}

Figure \ref{fig:qvalueevo} shows the time evolution of $Q$.
The $Q$ value is calculated from Equation (\ref{eq:qval}), where we calculate $\sigma_{x}$ from the root mean square of the radial velocity $\mathrm{d}x/\mathrm{d}t$.
Due to inelastic collisions between particles the kinetic energy decreases
 and the velocity dispersion and $Q$ decrease.
 When $Q$ becomes $Q \simeq 1\mbox{--}2$, the non-axisymmetric perturbations grow significantly \citep{Julian1966, Toomre1981}.
 Accordingly, the condition for the emergence of the self-gravity wakes is obtained from $Q\lesssim 2$
\citep{Salo1995,Ohtsuki2000,Daisaka2001} 
\begin{eqnarray}
 \tau \gtrsim \left\{ 
  \begin{array}{ll}
    0.08 \tilde r_\mathrm{H}^{-3}  & \mathrm{for} \,\,\, \tilde r_\mathrm{H} \lesssim 0.5 \\
    0.2 \tilde r_\mathrm{H}^{-3/2} & \mathrm{for} \,\,\, \tilde r_\mathrm{H} \gtrsim 0.5 \\
  \end{array} \right. .
 \label{eq:gicond}
\end{eqnarray}
When this condition is satisfied, the self-gravity wakes appear finally
regardless of $Q_\mathrm{ini}$ \citep{Salo2012}. 
As shown in Fig.\ref{fig:timeevo_of_snapshot}b, self-gravity wakes
 appear due to gravitational instability at $\tilde t=1.0$
 where  $\tilde t= t /( 2 \pi / \Omega ) $. 
Then the radial spacing of the self-gravity wakes is about
 $\lambda_\mathrm{cr}/2$ \citep{Salo2004}. 

After reaching about unity $Q$ increases with time and finally the
 time-averaged $Q$ becomes almost constant $\simeq 2$ after 5 orbital
 periods.   
In this state, the radial spacing is about $\lambda_\mathrm{cr}$
 \citep{Daisaka1999,Salo2004}.
Since heating by gravitational scattering and cooling by inelastic
 collisions are balanced, the time-averaged velocity dispersion is
 constant.
Figure \ref{fig:timeevo_of_snapshot}c shows the equilibrium state of the
 self-gravity wakes. 
 The high density region is trailing and inclined by $10 \mbox{--} 30^\circ$ with
 respect to the tangential direction. 

We performed simulations with various ring parameters and confirmed that
 the equilibrium state is reached after 5 orbital periods in all models  
 if the axisymmetric overstability does not exist.
Thus, hereafter we analyze the simulation result only for
$\tilde t= 5 \mbox{--} 20$. 

\subsection{Two-Dimensional Autocorrelation Function}

\begin{figure}
 \begin{minipage}{0.49\hsize}
 \begin{center} (a) $\tilde t = 5.0$ \end{center}
   \includegraphics[width=\columnwidth]{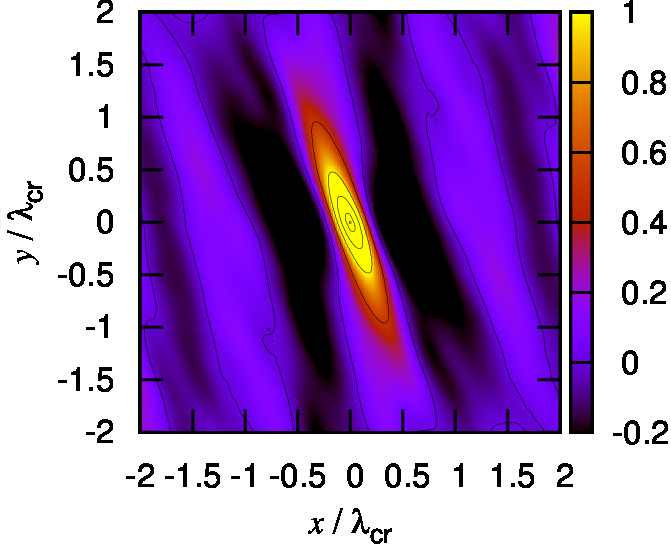}
 \end{minipage}
 \begin{minipage}{0.49\hsize}
   \begin{center} (b) $\tilde t = 10.0$ \end{center}
   \includegraphics[width=\columnwidth]{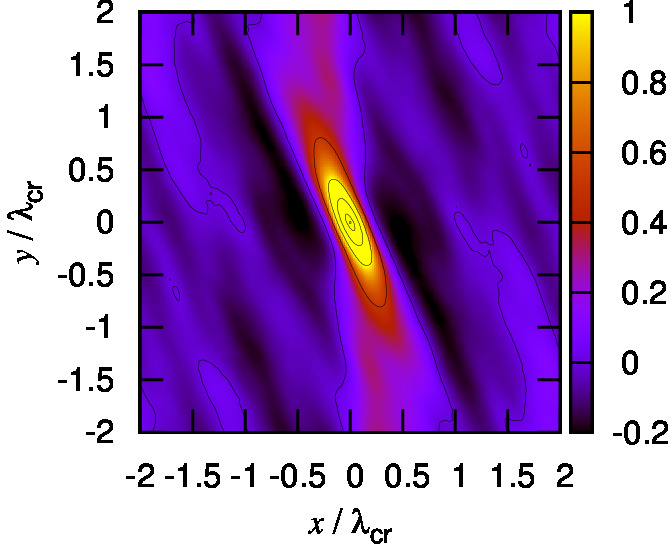}
 \end{minipage}
 \begin{minipage}{0.49\hsize}
 \begin{center} (c) $\tilde t = 15.0$ \end{center}
   \includegraphics[width=\columnwidth]{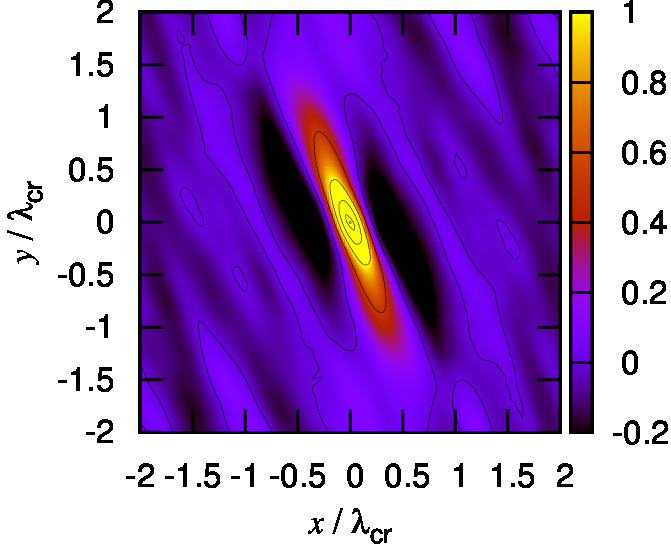}
 \end{minipage}
 \begin{minipage}{0.49\hsize}
 \begin{center} (d) Time Averaged \end{center}
   \includegraphics[width=\columnwidth]{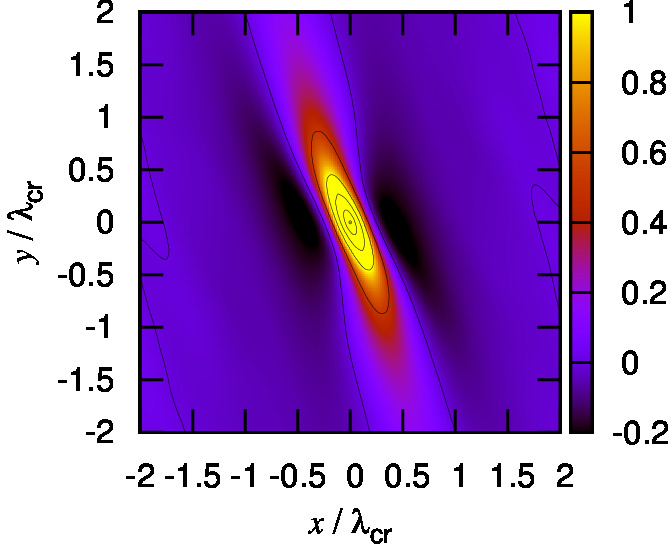}
 \end{minipage}
 \caption{Snapshots of the two-dimensional autocorrelation function  
  at $\tilde t=5.0$ (a), $10.0$ (b), and $15.0$ (c) for model 1.
  The panel (d) shows the time-averaged autocorrelation over
  $\tilde t=5.0 \mbox{--} 20.0$. 
}
\label{fig:timeevo_of_cor}
\end{figure}

Since the self-gravity wakes are a transient and recurrent feature we
 need to investigate them in a statistical way. 
Two-dimensional autocorrelation function has been widely used to analyze
 self-gravity wakes \citep{Toomre1991, Salo1995, Daisaka1999, Salo2004}.  
We use the two-dimensional autocorrelation function of the surface
 density given by
\begin{equation}
 \xi(x,y,t) = 
 \frac{1}{\Sigma_0^2 L^2} \int_{-L/2}^{L/2} \! 
 \int_{-L/2}^{L/2} \Sigma(x+x',y+y',t) \Sigma(x',y',t) 
 \mathrm{d}x' \mathrm{d}y' - 1.
\end{equation}
We calculate the autocorrelation with the uniform grid of $256 \times 256$.
We first calculate the surface density at each grid from particle distribution and then evaluate the integral.
Examples are shown in Figure \ref{fig:timeevo_of_cor}.
Though there are particle aggregates with various shapes in the surface
 density snapshot, the autocorrelation analysis enables us to extract a
 smooth wake structure.
As shown in Figure \ref{fig:timeevo_of_cor}a, b, and c, 
 although the wake structure is almost steady, it slightly fluctuates with time.
Therefore we introduce the time-averaged autocorrelation:
\begin{equation}
 \bar \xi(x,y) = \frac{1}{t_2-t_1} \int_{t_1}^{t_2} \xi(x,y,t') \mathrm{d}t',
\end{equation}
 where we set $\tilde t_1=5$ and $\tilde t_2 = 20$.
 We approximate the time integral by the sum of 150 snapshots in $\tilde t = 5.0 \mbox{--} 20.0$ with $\Delta \tilde t = 0.1$.
Figure \ref{fig:timeevo_of_cor}d shows the time-averaged
autocorrelation over $\tilde t=5.0 \mbox{--} 20.0$.
The small variation is completely smoothed out.
The inclined structure at the center corresponds to the structure of the
 high density region in the self-gravity wakes. 
This inclination means that the structure is trailing and the pitch
 angle is about $20^\circ$. 
 The regions with negative autocorrelation coefficient correspond to voids between self-gravity wakes.

It should be noted that the grid of $256\times 256$ is reasonable.
One might worry that the number of grids larger than that of
particles leads to large density error. This is always true for low
density parts where particles are sparsely distributed even if the
number of particles are larger than that of grids. However, it is not
problematic here since we focus on high density parts automatically
considering the autocorrelation function. In addition, time-averaging
of the autocorrelation function reduces the density error. To
demonstrate a suitable grid size Figure 4 demonstrates the validity of
the grid size. The function shape is correctly caught by the grid of
$256\times 256$. The typical discretization error of the pitch angle is
smaller than $\simeq 4$\%.

\begin{figure}
 \begin{minipage}{0.33\hsize}
 \begin{center} (a) $64\times 64$\end{center}
   \includegraphics[width=\columnwidth]{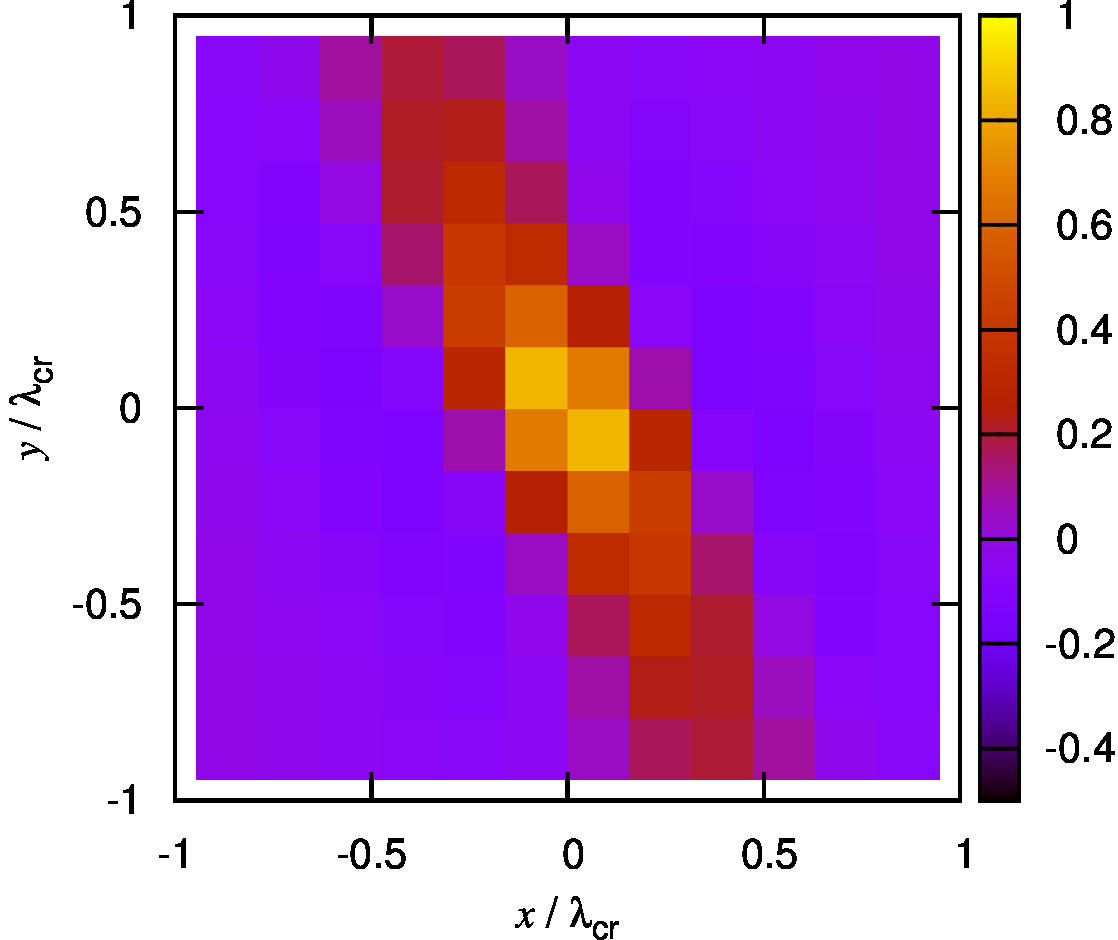}
 \end{minipage}
 \begin{minipage}{0.33\hsize}
 \begin{center} (b) $128\times 128$\end{center}
   \includegraphics[width=\columnwidth]{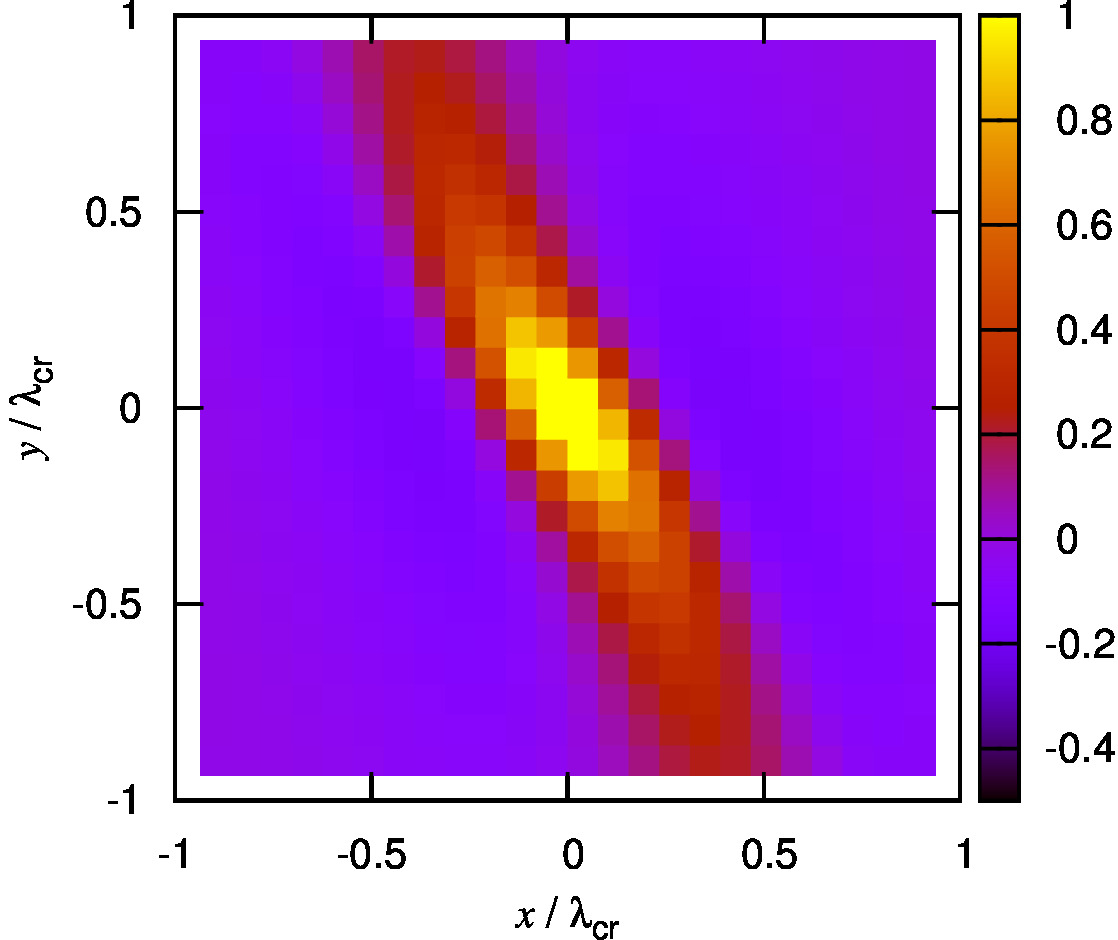}
 \end{minipage}
 \begin{minipage}{0.33\hsize}
 \begin{center} (c) $256\times 256$\end{center}
   \includegraphics[width=\columnwidth]{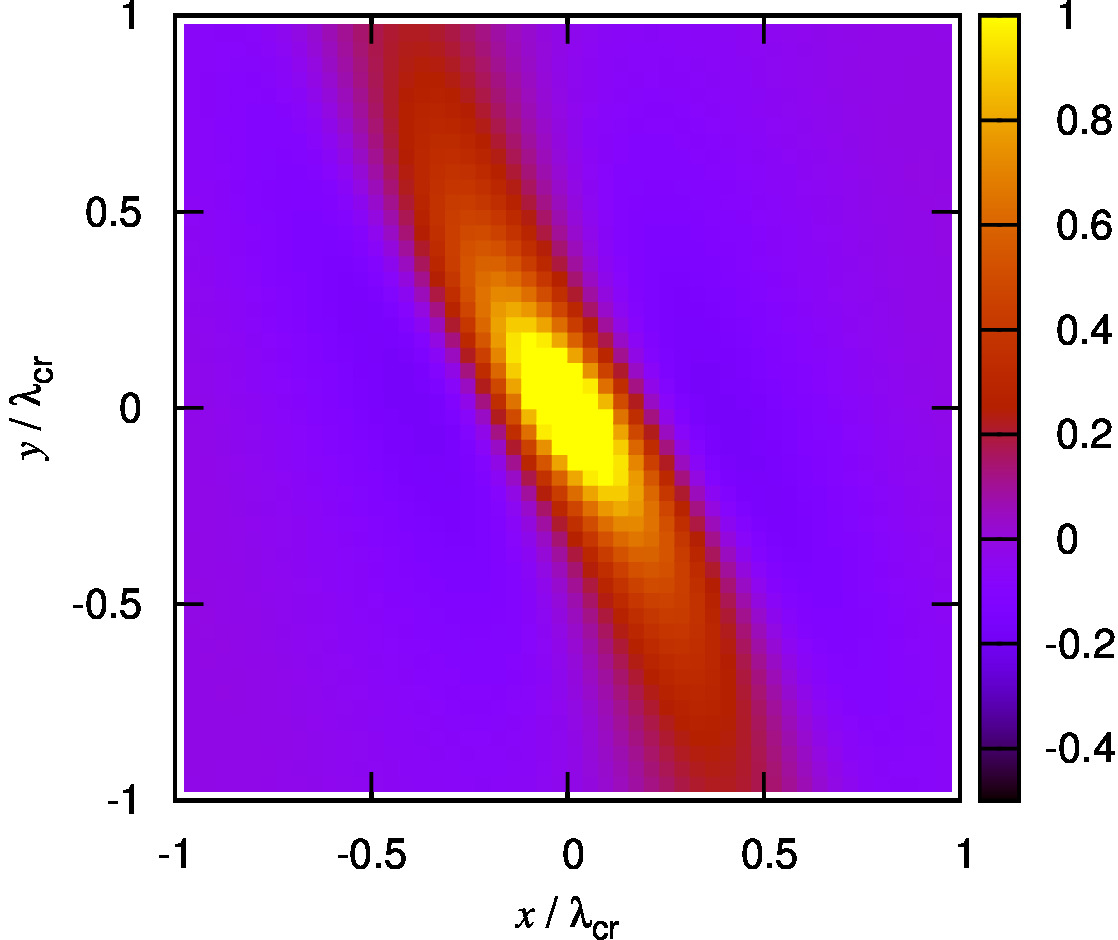}
 \end{minipage}
 \caption{Time-averaged autocorrelation functions for model 30 with grid of (a) $64\times 64$, (b) $128 \times 128$, and (c) $256\times256$.
}
\label{fig:bindep}
\end{figure}

\subsection{Calculation of Pitch Angle \label{sec:calcpitch}}
\begin{figure}
  \plottwo{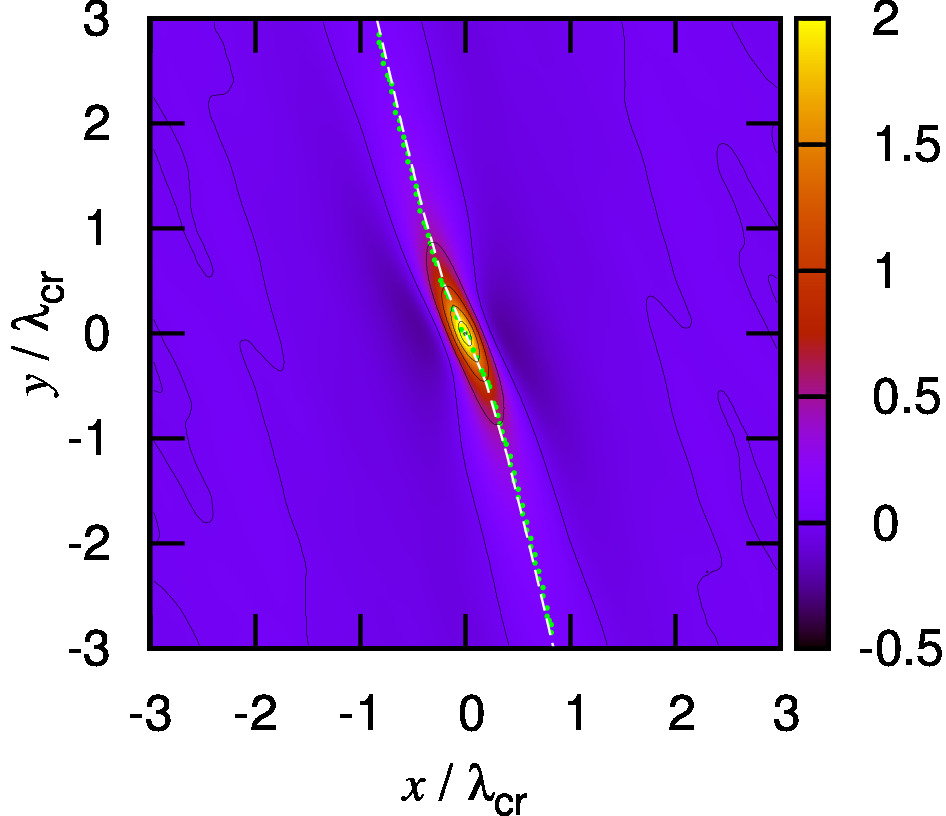}{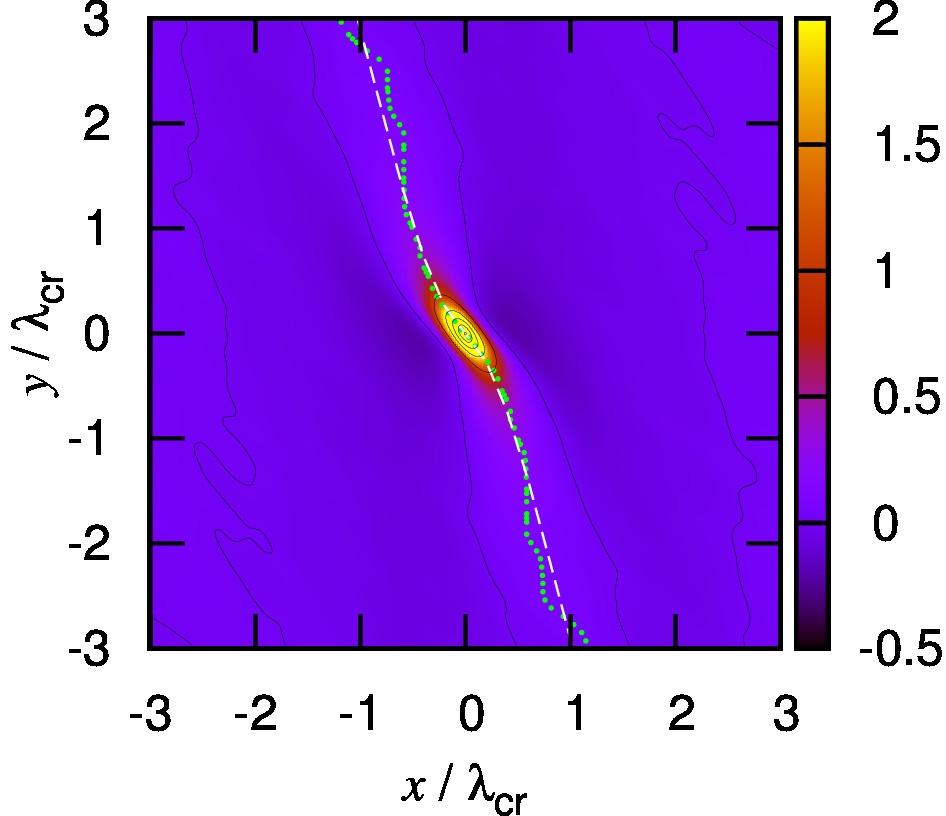}
 \caption{
   Autocorrelation function and the extracted ridge (dotted) for models 1 (left) and 42 (right). 
   The dashed curve shows the fitting function given by Equation (\ref{eq:fitfunc}) .
}
\label{fig:extracted_ridge}
\end{figure}

\begin{figure}
  \plotone{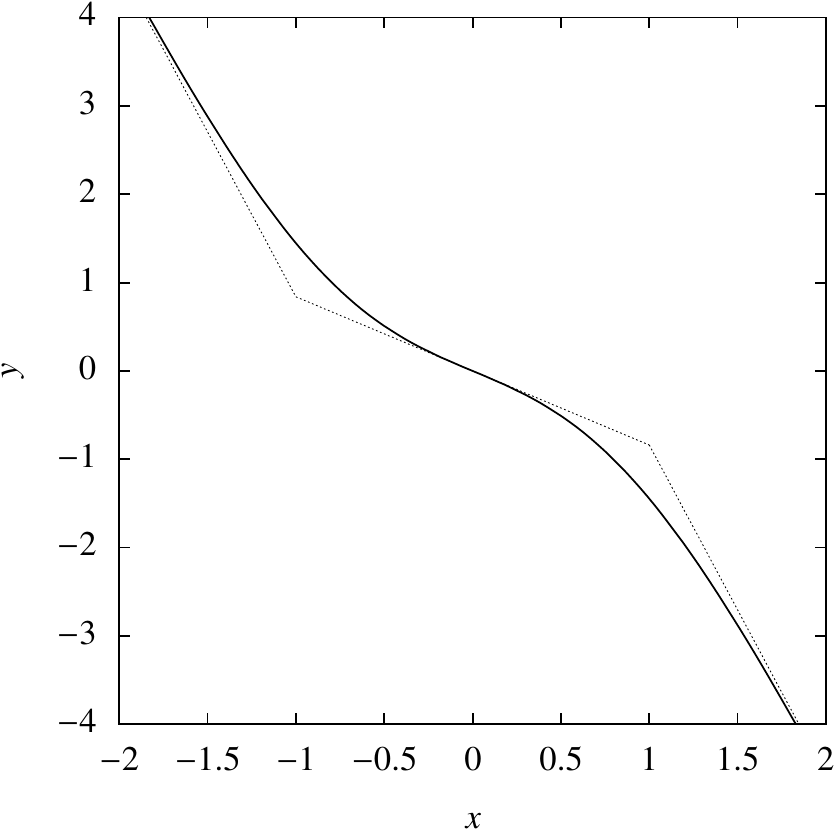}
  \caption{Fitting function for the ridge of autocorrelation with $\theta_\mathrm{i}=50^\circ$, $\theta_\mathrm{o}=20^\circ$, and $x_\mathrm{b}=1$ (solid line). The dotted line denotes the asymptotic lines the pitch angles of which are $\theta_\mathrm{i}$ and $\theta_\mathrm{b}$. The breakpoint is $x_\mathrm{b}$.
}
\label{fig:fit_sample}
\end{figure}

We calculate the pitch angle from the time-averaged autocorrelation.
 First, we extract the ridge of the dense
 part of autocorrelation by considering the integral curve of the
 gradient field of the autocorrelation function.
Figure \ref{fig:extracted_ridge} shows examples of the ridge of
 autocorrelation, which confirms the previous results that the slope of
 the ridge line varies with the distance from the center \citep{Salo2004}.
The slope in the outer part of the autocorrelation function is steeper than that in the inner part.
Clearly we cannot fit the ridge by one line.
Therefore we fit the ridge with the following function:
\begin{equation}
 y = - \cot \theta_\mathrm{o} x +
     (\cot \theta_\mathrm{o} - \cot \theta_\mathrm{i})
     x_\mathrm{b} \mathrm{erf} \left(\frac{\sqrt{\pi} x}{2x_\mathrm{b}} \right),
 \label{eq:fitfunc}
\end{equation}
where $\theta_\mathrm{i}$, $\theta_\mathrm{o}$, and $x_\mathrm{b}$ are parameters.
  The slope of the tangent to the fitting curve around the center $|x| \ll x_\mathrm{b}$ is $\cot \theta_\mathrm{i}$ and that for $|x|\gg x_\mathrm{b}$ is $\cot \theta_\mathrm{o}$.
  Hereafter, we define the inner and outer pitch angles as $\theta_\mathrm{i}$ and $\theta_\mathrm{o}$, where $\theta_\mathrm{i}$ and $\theta_\mathrm{o}$ correspond to the pitch angles in the inner and outer parts of the autocorrelation function, respectively.
  The value $x_\mathrm{b}$ means the typical length where the pitch angle changes from $\theta_\mathrm{i}$ to $\theta_\mathrm{o}$.
  An example of the fitting function is shown in Figure \ref{fig:fit_sample}.

We introduce a line integral of autocorrelation function along the ridge
 defined by Equation (\ref{eq:fitfunc}) as
\begin{equation}
 S(\theta_\mathrm{i}, \theta_\mathrm{o}, x_\mathrm{b}) =
 \int \bar \xi(x, y) \mathrm{d} \mbox{\boldmath $l$}.
 \label{eq:lineint}
\end{equation}
We calculate the integral by Simpson's rule and find $\theta_\mathrm{i}$,
 $\theta_\mathrm{o}$, and $x_\mathrm{b}$ that maximize $S$ by the downhill simplex method \citep{Nelder1965, Press2007}.

We obtain the representative pitch angles from the time averaged autocorrelation function.
However, the actual pitch angles fluctuate with time around the representative values.
Thus, in order to estimate the fluctuation degree, we calculate the median absolute deviation from the time series of the inner and outer pitch angles calculated from the autocorrelation functions for $\tilde t=5.0 \mbox{--} 20.0$.
Examples of the optimized parameters and the median absolute deviation are
$\theta_\mathrm{i}=(25.8 \pm 4.1)^\circ$, $\theta_\mathrm{o}=(13.3 \pm 4.2) ^\circ,$ and
 $x_\mathrm{b}/\lambda_\mathrm{cr}=0.26$ for model 1 and
 $\theta_\mathrm{i}=(41.9 \pm 7.2)^\circ$, $\theta_\mathrm{o}=(14.7 \pm 6.8)^\circ,$ and
 $x_\mathrm{b}/\lambda_\mathrm{cr}=0.34$ for model 42. 
As shown in Figure \ref{fig:extracted_ridge}, the ridge curves agree well with Equation (\ref{eq:fitfunc}).
Model 42 shows the deviation of the extracted ridge from the fitting curve in the low autocorrelation region.
This is because that the ridge extracted by the gradient field is sensitively affected by the noise.
In the low autocorrelation region, the fitting curve is more robustly determined than the extracted ridge.

The inner and outer pitch angles $\theta_\mathrm{i}$ and $\theta_\mathrm{o}$, and
the breakpoint position $(x_\mathrm{b}, y_\mathrm{b})$ are summarized in Tables \ref{tbl:model1} and \ref{tbl:model2} 
where $y_\mathrm{b}$ is calculated from $y_\mathrm{b} = -x_\mathrm{b} / \tan \theta_\mathrm{i}$.
The breakpoint $x_\mathrm{b}$ is generally less than $\lambda_\mathrm{cr}$ and about $(0.1 \mbox{--}0.5)\lambda_\mathrm{cr}$.


\section{Parameter Dependence}

We first adopt the identical particle model and examine the dependence
 of the pitch angle on the restitution coefficient $\epsilon$, the optical
 depth $\tau$, and the Hill radius of a particle pair relative to the sum of their physical radii $\tilde r_\mathrm{H}$.  
Next, we investigate the effect of the size distribution of ring
 particles. 

\subsection{Restitution Coefficient \label{seceps}}

\begin{figure}
  \plotone{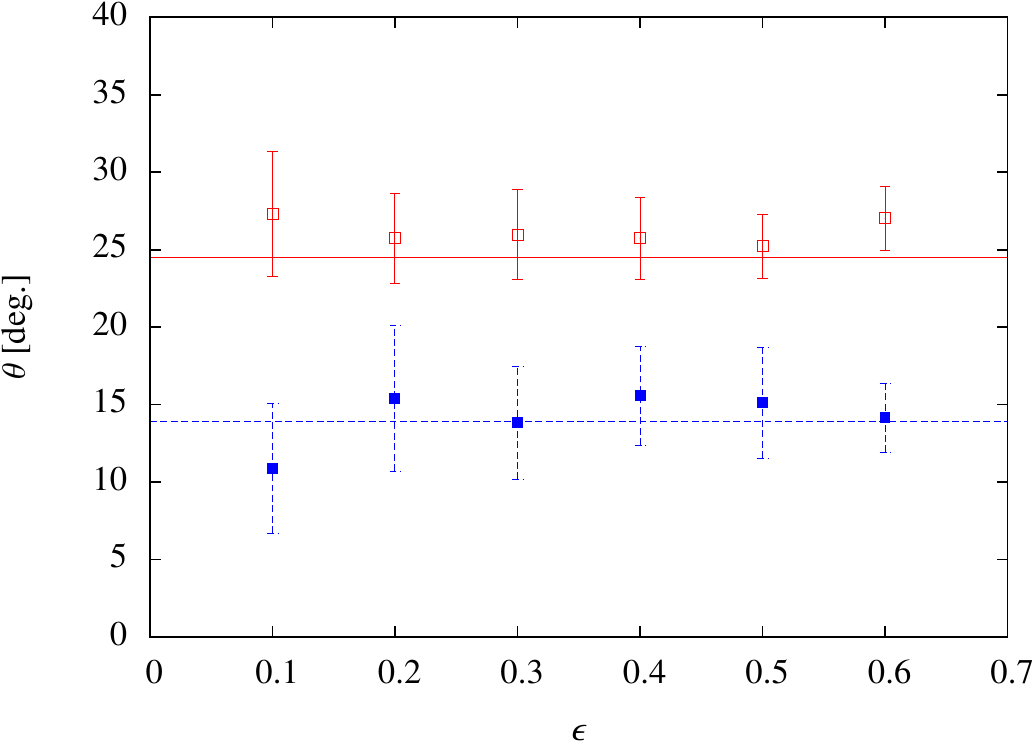}
  \caption{
 Inner and outer pitch angles $\theta_\mathrm{i}$ and
 $\theta_\mathrm{o}$ are plotted against the restitution coefficient
 $\epsilon$ with $\tau=0.7$ and 
 $\tilde r_\mathrm{H}=0.81$ (models 2--7).  
 The open and filled symbols denote $\theta_\mathrm{i}$ and
 $\theta_\mathrm{o}$, respectively. 
 The solid and dashed lines denote $\theta_\mathrm{i}$ and
 $\theta_\mathrm{o}$ for the velocity-dependent restitution coefficient
 model, respectively (model 8). 
}
\label{fig:pitch_edep}
\end{figure}

First we adopt the constant restitution coefficient model and
 investigate its effect on the pitch angle.
The restitution coefficients adopted here are 
 $\epsilon = 0.1, 0.2, 0.3, 0.4, 0.5$ and $0.6$ (models 2--7). 
The other parameters are $\tau = 0.7$ and $\tilde r_\mathrm{H} = 0.81$.
The results are summarized in Figure \ref{fig:pitch_edep}.
In all models, the inner and outer pitch angles barely depend on $\epsilon$, and their mean values are about $26.1^\circ$ and $14.1^\circ$, respectively.
As shown by \cite{Salo2004}, the inner pitch angle is larger than the outer one. 
Since the influence of the self-gravity relative to the shear is strong
 around the center of self-gravity wakes, the shape in the inner part of
 the autocorrelation function is nearly spherical. 
Conversely, the shape in the outer part is elongated due to the shear. 
Thus, we have $\theta_\mathrm{i} > \theta_\mathrm{o} $.

It is expected that the pitch angle does not change for the
 velocity-dependent restitution coefficient model, since the pitch angles barely depend on $\epsilon$ in the constant restitution coefficient
 models. 
This is confirmed by the simulation with the velocity-dependent
 restitution coefficient (model 8) where we assume 
 $a=1.25 \times 10^5\, \mathrm{km}$, 
 $\rho_\mathrm{p}=0.45 \, \mathrm{g} \, \mathrm{cm}^{-3}$, 
 and $R_\mathrm{p} = 120 \, \mathrm{cm}$ (Fig.\ref{fig:pitch_edep}). 
The inner and outer pitch angles are $24.5 ^\circ$ and $13.9 ^\circ$, respectively.
They are approximately the same as those for the constant restitution coefficient models. 
It is clearly shown that the pitch angle is independent of the restitution coefficient.  

\subsection{Optical Depth \label{optdep}}

\begin{figure}
 \plottwo {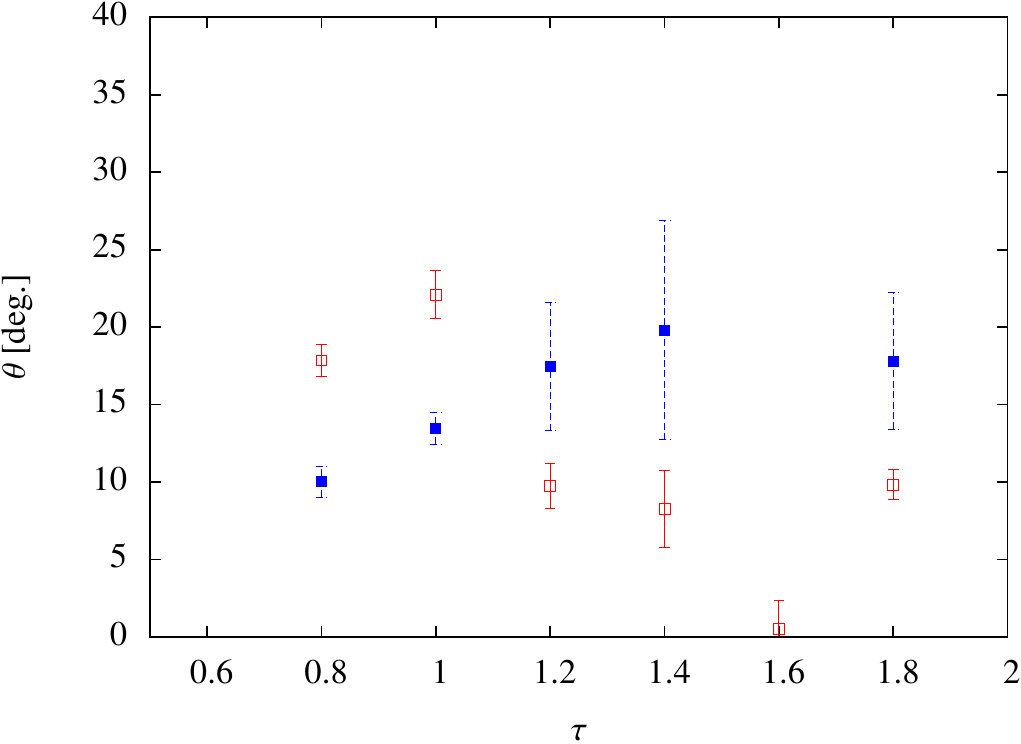} {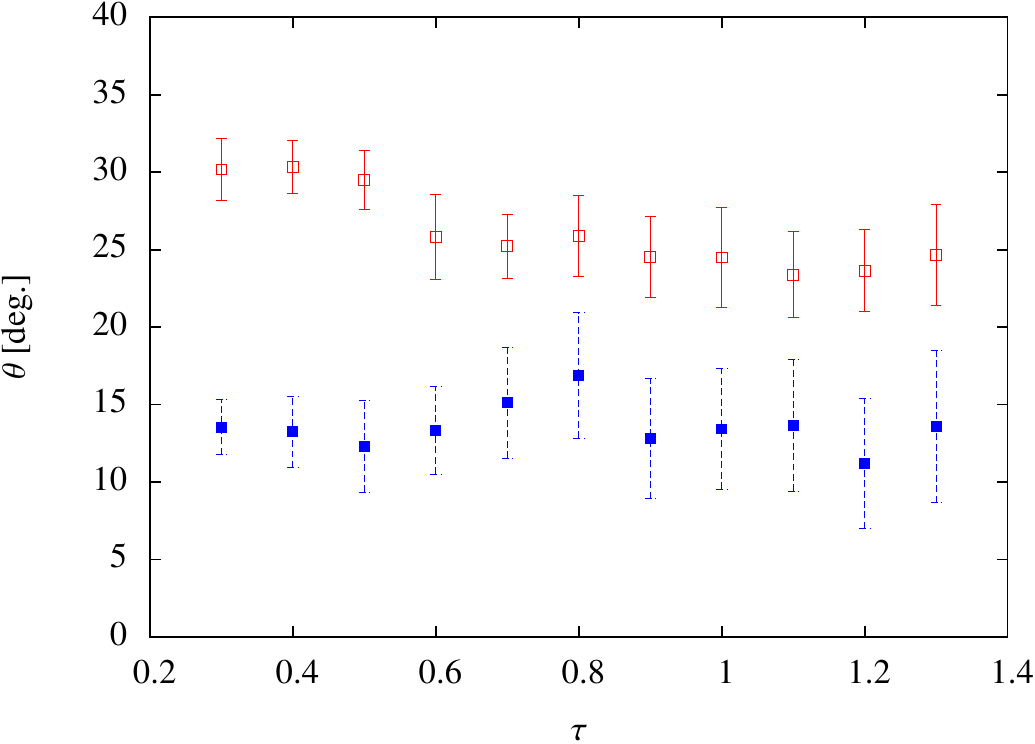}
 \caption{
 Inner and outer pitch angles $\theta_\mathrm{i}$ (open) and
 $\theta_\mathrm{o}$ (filled) are plotted against the optical depth $\tau$. 
 The restitution coefficient is $\epsilon = 0.5$ and 
 Hill radius of a particle pair relative to the sum of their physical radii
 is $\tilde r_\mathrm{H}=0.6$ (left, models 9--14)
 and $0.81$ (right, models 15--24).
 }
\label{fig:pitch_tdep}
\end{figure}

\begin{figure}
 \plottwo{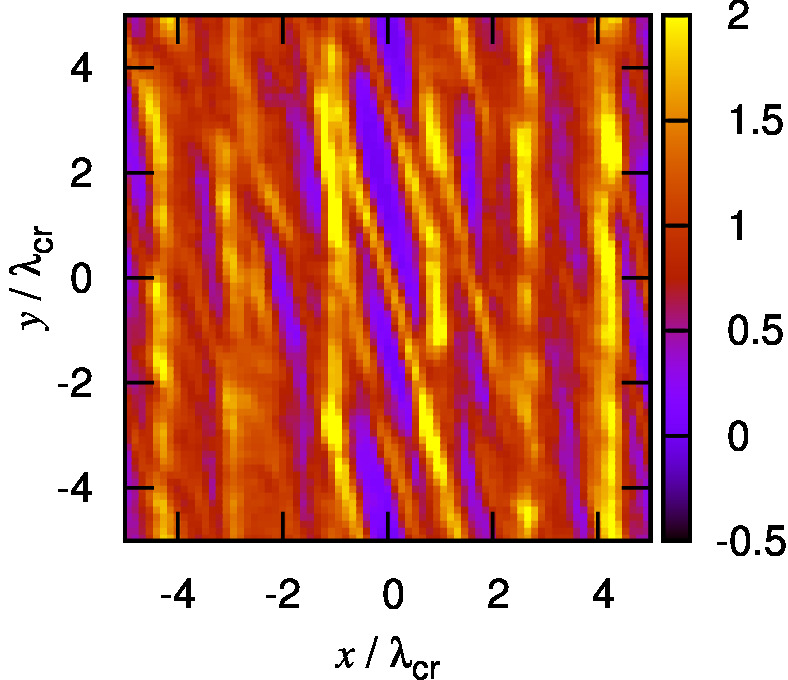}{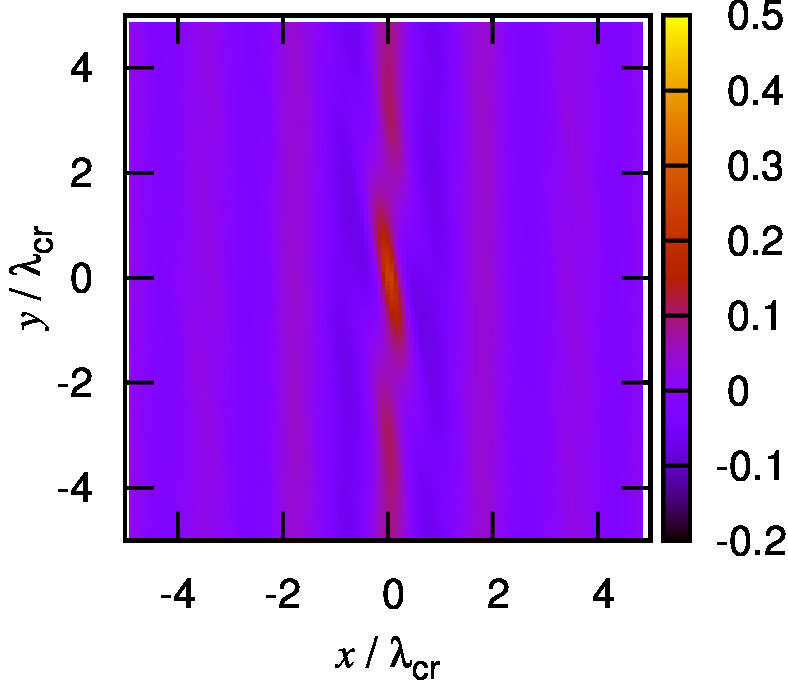}
 \caption{
 Snapshots of the surface density (left) and its autocorrelation (right)
 for the model with $\tau = 1.6$, $\tilde r_\mathrm{H} = 0.6$, and
 $\epsilon = 0.5$ (model 15) at $\tilde t = 19.0$. 
 } 
\label{fig:os_dencor}
\end{figure}

We investigate the dependence of the pitch angle on the dynamical
 optical depth $\tau$ with fixed $\tilde r_\mathrm{H}$ and $\epsilon$.  
The restitution coefficient is $\epsilon = 0.5$ and the Hill radius of a particle pair relative to the sum of their physical radii is $\tilde r_\mathrm{H} = 0.6$ and $0.81$.  
 The optical depths are $\tau = 0.8 \mbox{--} 1.8$ for 
 $\tilde r_\mathrm{H} =0.6$ (models 9--14) and 
 $\tau = 0.3 \mbox{--} 1.3$ for $\tilde r_\mathrm{H} = 0.81$ 
 (models 15--24). 
The results are shown in Figure \ref{fig:pitch_tdep}.
For $\tilde r_\mathrm{H}=0.6$ the inner pitch angle roughly decreases with $\tau$, and
 becomes lower than $10^\circ$. For $\tau>1.0$ the outer pitch angle is larger than the inner one.
However, it is arguable whether these values can be
 considered as the representative pitch angle of self-gravity wakes. 
Figure \ref{fig:os_dencor} illustrates the surface density and its autocorrelation for $\tau = 1.6$ and $\tilde r_\mathrm{H} = 0.6$ (model 13) at $\tilde t = 19$. 
We find that the axisymmetric structure and the non-axisymmetric
 self-gravity wakes coexist.
This axisymmetric structure is caused by the overstable oscillation
 \citep{Schmit1995, Schmidt2001, Salo2001, Schmidt2003, Latter2006a}. 
The axisymmetric structure appears for $\tau \gtrsim 1$ and 
 $\tilde r_\mathrm{H} \sim 0.6$. 
The overstable oscillation causes the small pitch angles \citep{French2007}. 

\begin{figure}
 \begin{minipage}{0.49\hsize}
 \begin{center} (a) $\tau = 0.30$ (model 15) \end{center}
   \includegraphics[width=\columnwidth]{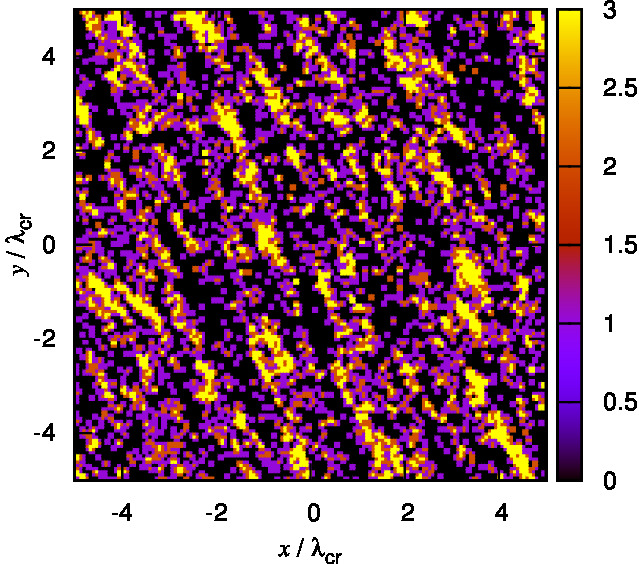}
 \end{minipage}
 \begin{minipage}{0.49\hsize}
   \begin{center} (b) $\tau = 0.40$ (model 16) \end{center}
   \includegraphics[width=\columnwidth]{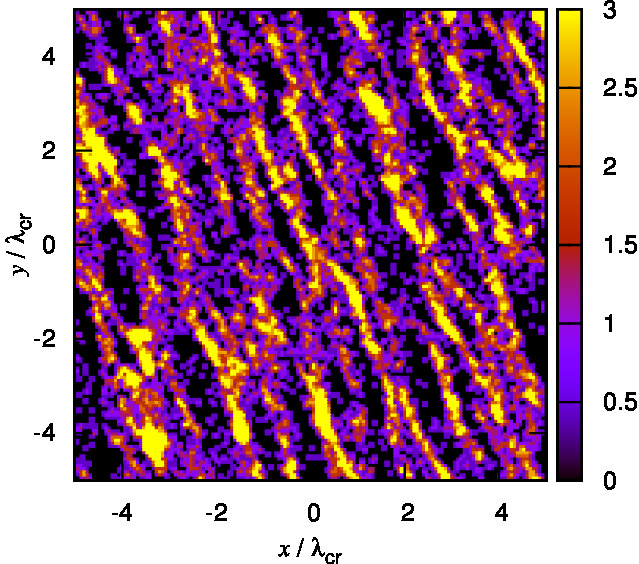}
 \end{minipage}
 \begin{minipage}{0.49\hsize}
   \begin{center} (c) $\tau = 0.50$ (model 17) \end{center}
   \includegraphics[width=\columnwidth]{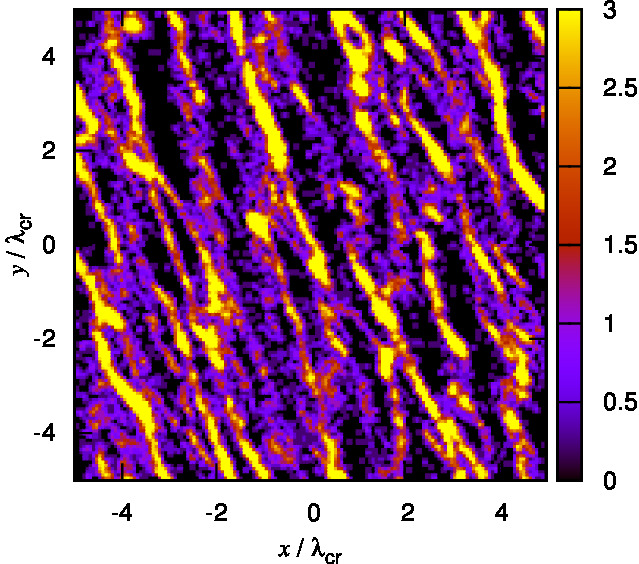}
 \end{minipage}
 \begin{minipage}{0.49\hsize}
   \begin{center} (d) $\tau = 0.60$ (model 18) \end{center}
   \includegraphics[width=\columnwidth]{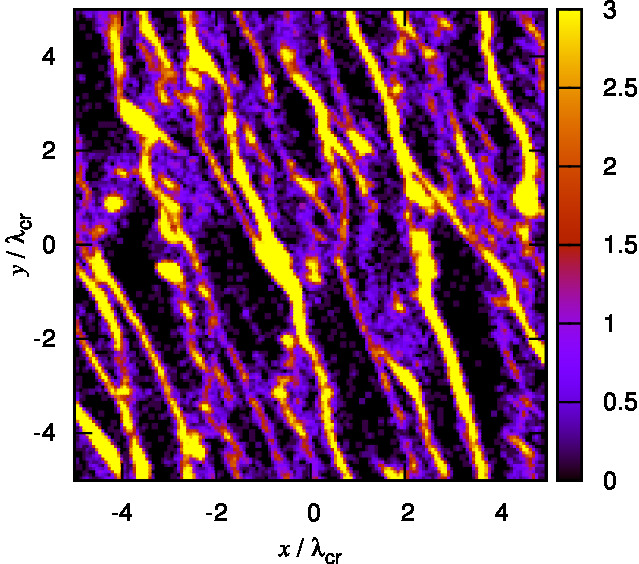}
 \end{minipage}
 \caption{
   Snapshots of the surface density normalized by the mean surface density at $\tilde t=20.0$ for models 15 (a), 16 (b), 17 (c), and 18 (d). 
}
\label{fig:snaprh081}
\end{figure}

On the other hand, for $\tilde r_\mathrm{H} = 0.81$ the outer pitch
 angle does not depend on $\tau$, and for $\tau > 0.5$ the inner pitch
 angle does not have clear dependence on $\tau$.
For $\tau > 0.5$ the mean inner and outer pitch angles are
 $24.7^\circ$ and $13.7^\circ$, respectively.
For $0.3 \leq \tau \leq 0.5$ the inner pitch angle is systematically
 larger than $24.7^\circ$.
The inner pitch angles are $30.2^\circ$, $30.3^\circ$, and $29.5^\circ$
 for $\tau=0.3$ (model 15), $0.4$ (model 16), and $0.5$ (model
 17), respectively.
Figure \ref{fig:snaprh081} represents the surface density snapshots at
 $\tilde t=20$ for $\tau=0.30$ (model 15), $0.40$ (model 16),
 $0.50$ (model 17), and $0.60$ (model 18).
From Equation (\ref{eq:gicond}), the condition for self-gravity wake
 formation is $\tau \gtrsim 0.27$.
Thus the ring of model 15 is marginally unstable and the self-gravity
 wakes are small and faint.
As $\tau$ increases (models 16, 17, and 18), the clearer long self-gravity
 wakes and voids develop.
A possible explanation for the large inner pitch angle at small $\tau$
 is as follows.
In marginally unstable rings, the self-gravity is weak and the resulting self-gravity wakes are faint.
Since the wake clumps are loosely bound, they are easily destroyed by the interactions among clumps before they are elongated.
In short we find that as long as the overstable oscillation does not
 occur and the clear self-gravity wakes form with large $\tau$, the
 inner pitch angle is almost independent of $\tau$.

\subsection{Hill Radius Relative to Sum of Particles Radii \label{secrh}}
\begin{figure}
 \begin{minipage}{0.33\hsize}
  \begin{center}
   \includegraphics[width=\textwidth]{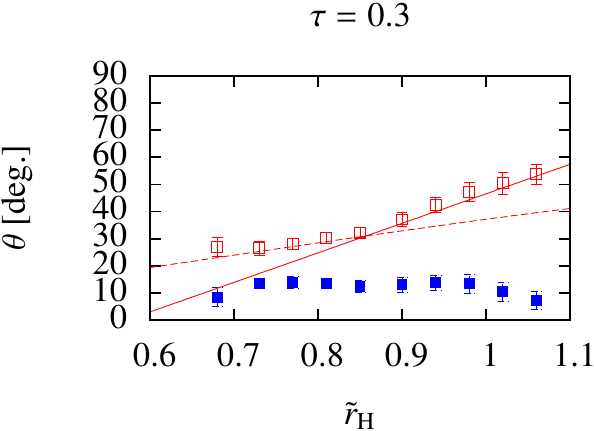}
  \end{center}
 \end{minipage}
 \begin{minipage}{0.33\hsize}
  \begin{center}
   \includegraphics[width=\textwidth]{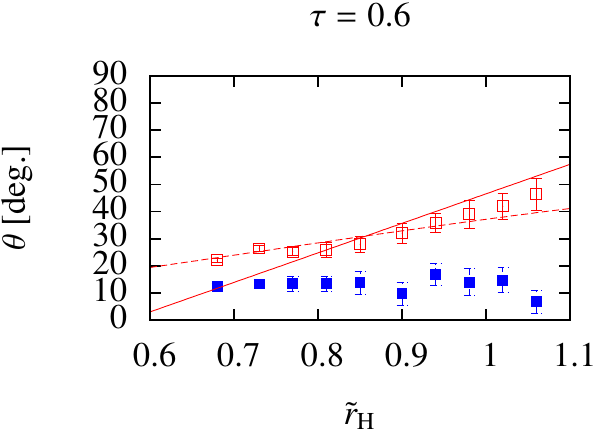}
  \end{center}
 \end{minipage}
 \begin{minipage}{0.33\hsize}
  \begin{center}
   \includegraphics[width=\textwidth]{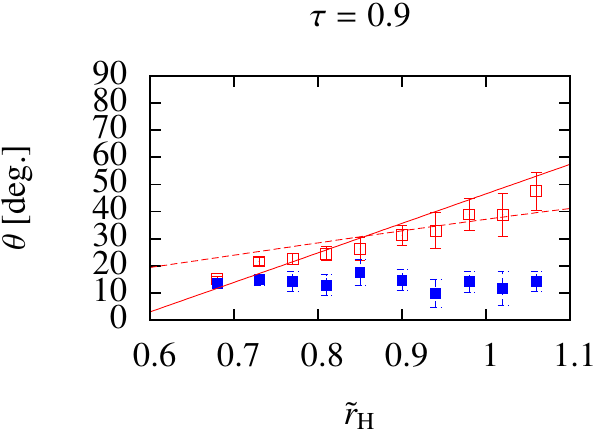}
  \end{center}
 \end{minipage}
 \caption{
 Inner and outer pitch angles $\theta_\mathrm{i}$ (open) and
 $\theta_\mathrm{o}$ (filled) plotted against $\tilde r_\mathrm{H}$ with the optical depth $\tau = 0.3$ (left,
 models 25--34), $\tau = 0.6$ (middle, models 35--43), and $\tau = 0.9$
 (right, models 44--53).  
The solid lines denote the fitting formula given by Equations (\ref{eq:pitchinner}), and 
the dashed lines denote the estimation given by Equation (\ref{eq:theory}) with $C_\mathrm{w}=0.45$ and $C_\mathrm{d}=1.0$. 
}
\label{fig:pitch_rdep}
\end{figure}

The dependence of the pitch angle on $\tilde r_\mathrm{H}$ is shown for $\tau = 0.3$ (models 25--34), $0.6$ (models 35--43), and $0.9$ (models 44--53) in Figure \ref{fig:pitch_rdep}.
 The outer pitch angle is independent of both $\tau$ and
 $\tilde r_\mathrm{H}$.
The mean outer pitch angles for $\tau=0.3, 0.6,$ and $0.9$ are $12.0^\circ$, $12.8^\circ$, and $13.8^\circ$, respectively.
This reason will be discussed in Section \ref{sec:disgal}.

On the other hand, the inner pitch angle increases with
 $\tilde r_\mathrm{H}$ from $30^\circ$ to $50^\circ$ for
 $\tilde r_\mathrm{H}>0.8$.
 The similar tendency was shown in the previous numerical studies \citep{Salo1995, Takeda2001}.
However, for $\tilde r_\mathrm{H}<0.8$ and $\tau=0.6$ (models 35, 36, and
 37), the inner pitch angle is almost constant, and for
 $\tilde r_\mathrm{H}=0.77$ and $\tau=0.3$ (model 27) the inner pitch
 angle is unexpectedly large.
The rings of these models are marginally unstable.
Thus, as discussed in Section \ref{optdep}, the pitch angle becomes large.

For $0.8 < \tilde r_\mathrm{H}$ , using the least-square fit method, we
 obtain the fitting formula of the inner pitch angle:
\begin{equation}
 \theta_\mathrm{i} \simeq
 \left\{
 \begin{array}{ccccc}
   (117.9 \pm 12.0) \tilde r_\mathrm{H} & + &
   (-65.1 \pm 11.3)  & \mathrm{deg}. & (\tau=0.3), \\
   (94.8 \pm 4.9) \tilde r_\mathrm{H} & + &
   (-52.0 \pm 4.6) & \mathrm{deg}. & (\tau=0.6), \\
   (113.3 \pm 21.6) \tilde r_\mathrm{H} & + &
   (-69.5 \pm 20.4) & \mathrm{deg}. & (\tau=0.9),
 \end{array}
 \right.
\end{equation}
 which shows that the dependence of $\theta_\mathrm{i}$ on
 $\tilde r_\mathrm{H}$ is independent of $\tau$.
Thus, fitting the results of all $\tau$ models we obtain a general
 formula for $\tilde r_\mathrm{H} > 0.8$ and $\tau=0.3$--$0.9$:
\begin{equation}
 \theta_\mathrm{i} \simeq
 (108.7 \pm 13.4) \tilde r_\mathrm{H} + (-62.2 \pm 12.8)\ \mathrm{deg}. 
 \label{eq:pitchinner}
\end{equation}

\subsection{Size Distribution \label{secsize}}

\begin{figure}
 \plotone{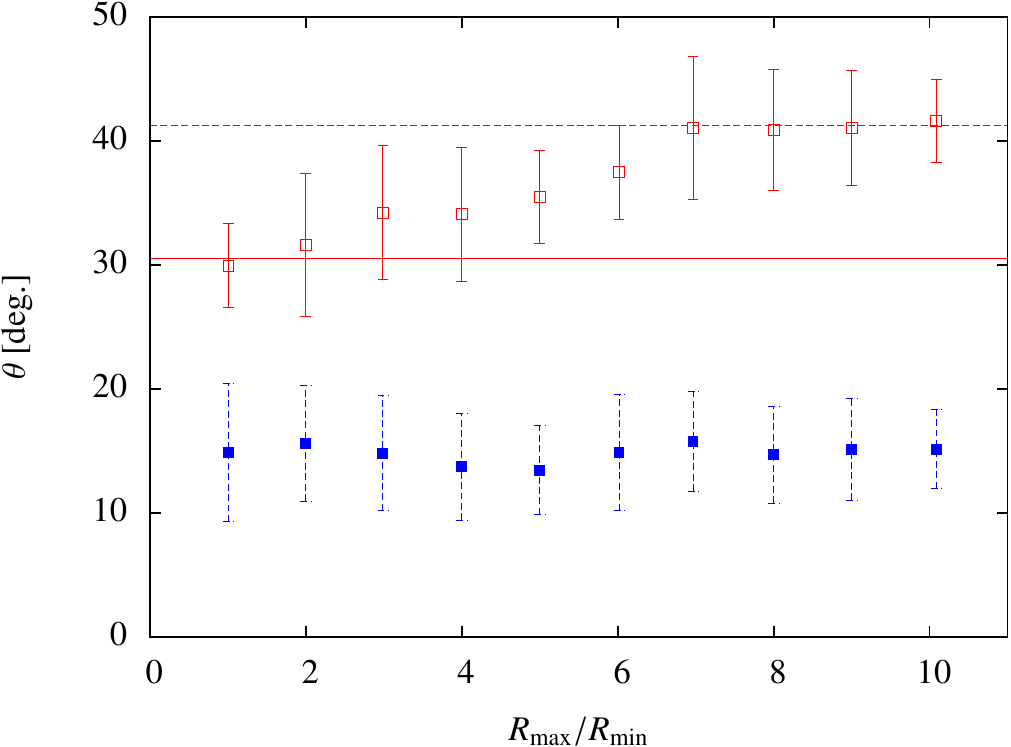}
 \caption{
 Inner and outer pitch angles $\theta_\mathrm{i}$ (open) and
 $\theta_\mathrm{o}$ (filled) plotted against
 $R_\mathrm{max}/R_\mathrm{min}$ for models 54--63.    
 The surface density, the Saturnicentric distance and the particle density are $\Sigma_0 = 50 \, \mathrm{g} \, \mathrm{cm}^{-2}$, 
 $a = 1.3 \times 10^5 \mathrm{km}$ and 
 $\rho_\mathrm{p} = 0.45 \, \mathrm{g} \, \mathrm{cm}^{-3}$, respectively. 
 The solid and dashed lines denote the estimation given by Equation (\ref{eq:theory}) with $C_\mathrm{w}=0.45$ (solid) and $C_\mathrm{w}=1.0$ (dashed), respectively. 
 \label{fig:pitch_rdep_size}
 }
\end{figure}

We investigate the effect of the size distribution of ring particles on the pitch angle.
We adopt the power-law size distribution described by Equation (\ref{eq:size}) with the lower and upper size limits (models 54--63).
Figure \ref{fig:pitch_rdep_size} illustrates the dependence of the pitch angle on the ratio $R_\mathrm{max}/R_\mathrm{min}$. 
The surface density, the Saturnicentric distance, and the particle density are $\Sigma_0=50 \, \mathrm{g} \, \mathrm{cm}^{-2}$, $a = 1.3 \times 10^5 \mathrm{km}$, and $\rho_\mathrm{p} = 0.45 \, \mathrm{g} \, \mathrm{cm}^{-3}$, respectively. 
The inner pitch angle increases with $R_\mathrm{max}/R_\mathrm{min}$.
The inner pitch angle with the size distribution is $30.0^\circ$--$41.6^\circ$ and is $39$\% larger than that for identical particles.
This is consistent with the previous results by \cite{Salo2004}.
As $R_\mathrm{max}/R_\mathrm{min}$ increases, the filling factor and density of self-gravity wakes increase. 
The higher density of self-gravity wakes leads to the larger pitch angle because the shear in the higher density wakes is weaker relative to the self-gravity. 
The details of this dependence will be discussed in Section
 \ref{sec:dest}. 
On the other hand, the outer pitch angle barely depends on $R_\mathrm{max}/R_\mathrm{min}$. The mean outer pitch angle is $14.8^\circ$.

\begin{figure}
 \plottwo{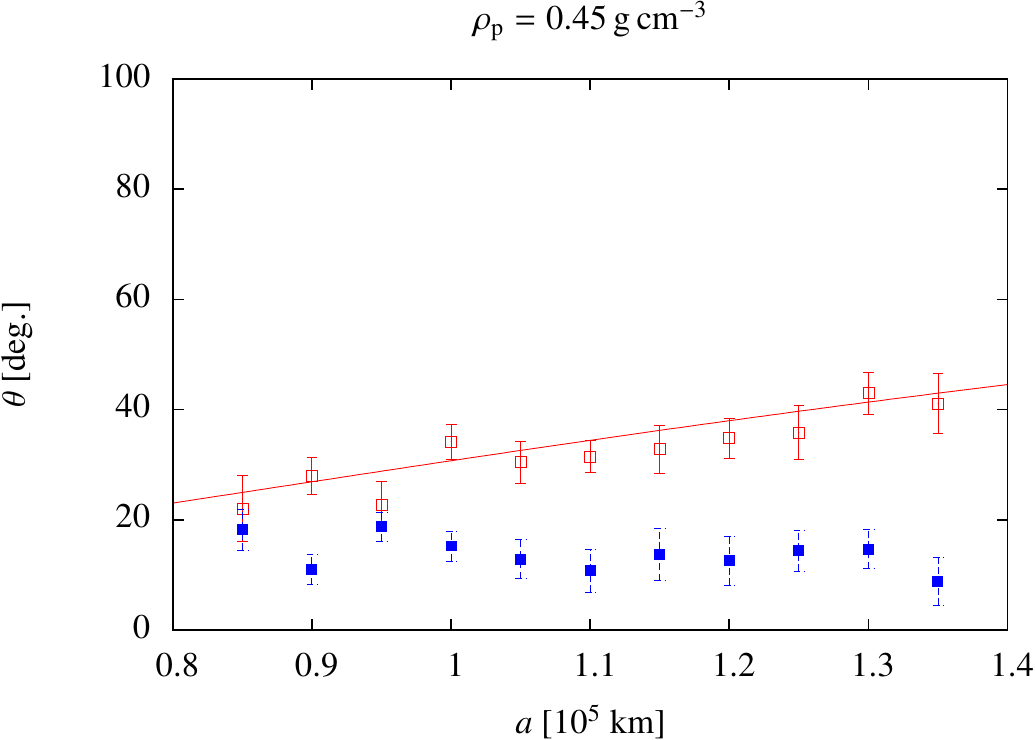}{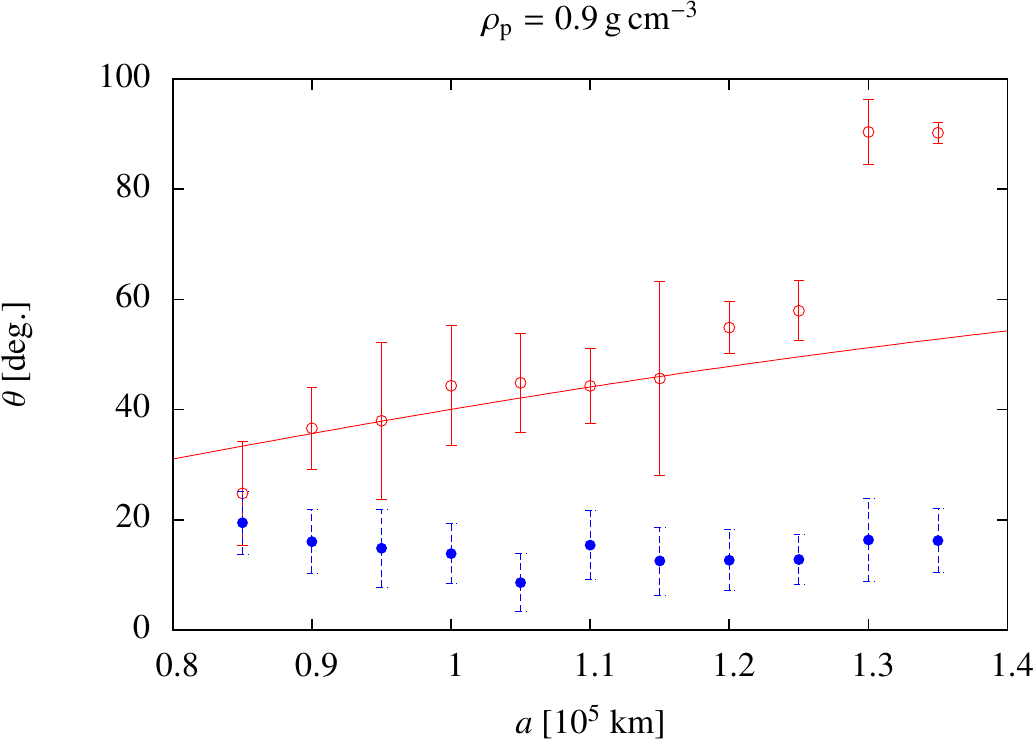}
 \caption{
 Inner and outer pitch angles $\theta_\mathrm{i}$ (open) and
 $\theta_\mathrm{o}$ 
 (filled) plotted against the Saturnicentric distance.
 The particle densities are 
 $\rho_\mathrm{p}=0.45 \, \mathrm{g} \, \mathrm{cm}^{-3}$ 
 (left, models 63--73) and
 $\rho_\mathrm{p}=0.45 \, \mathrm{g} \, \mathrm{cm}^{-3}$ 
 (right, models 74--84). 
 The solid lines denote the estimation given by Equation (\ref{eq:theory}) with $C_\mathrm{w}=1.0$  and $C_\mathrm{d}=1.0$. 
 \label{fig:pitch_sizea}
 }
\end{figure}

In Section \ref{secrh}, we found that the pitch angle for identical particles increases with $\tilde r_\mathrm{H}$.  
We expect the same trend for the pitch angle in the size distribution models. 
Figure \ref{fig:pitch_sizea} shows the dependence of the pitch angle on the Saturnicentric distance. 
The surface density is $\Sigma_0=50 \, \mathrm{g} \, \mathrm{cm}^{-2}$ and the particle density is 
$\rho_\mathrm{p} = 0.45 \, \mathrm{g} \, \mathrm{cm}^{-3}$ (models 63--73) and 
 $0.9 \, \mathrm{g} \, \mathrm{cm}^{-3}$ (models 74--84).
The lower and upper size limits are fixes as $R_\mathrm{min} = 41.7 \, \mathrm{cm}$ and $R_\mathrm{max} = 417 \, \mathrm{cm}$. 
The inner pitch angles for 
 $\rho_\mathrm{p} = 0.45 \, \mathrm{g} \, \mathrm{cm}^{-3}$ are smaller
 than those for $\rho_\mathrm{p} = 0.9 \, \mathrm{g} \, \mathrm{cm}^{-3}$. 
This is because smaller $\rho_\mathrm{p}$ corresponds to smaller $\tilde r_\mathrm{H}$. 
As expected, the inner pitch angle generally increases with the Saturnicentric distance. 
For $\rho_\mathrm{p}=0.45 \, \mathrm{g} \, \mathrm{cm}^{-3}$ the pitch angle increases with the Saturnicentric distance $a$ 
from $22.0^\circ$ to $43.0^\circ$, and the outer pitch angle has no trend and ranges within $8.8 ^\circ \mbox{--} 18.8 ^\circ$.
Similarly, for $\rho_\mathrm{p}=0.9 \, \mathrm{g} \, \mathrm{cm}^{-3}$, the pitch angle increases with $a$ from $24.8^\circ$ to $58.0^\circ$ for $a \le 1.25 \times 10^5 \, \mathrm{km}$, 
and the outer pitch angle is $8.6 ^\circ \mbox{--} 19.4 ^\circ$.

\begin{figure}
  \plottwo{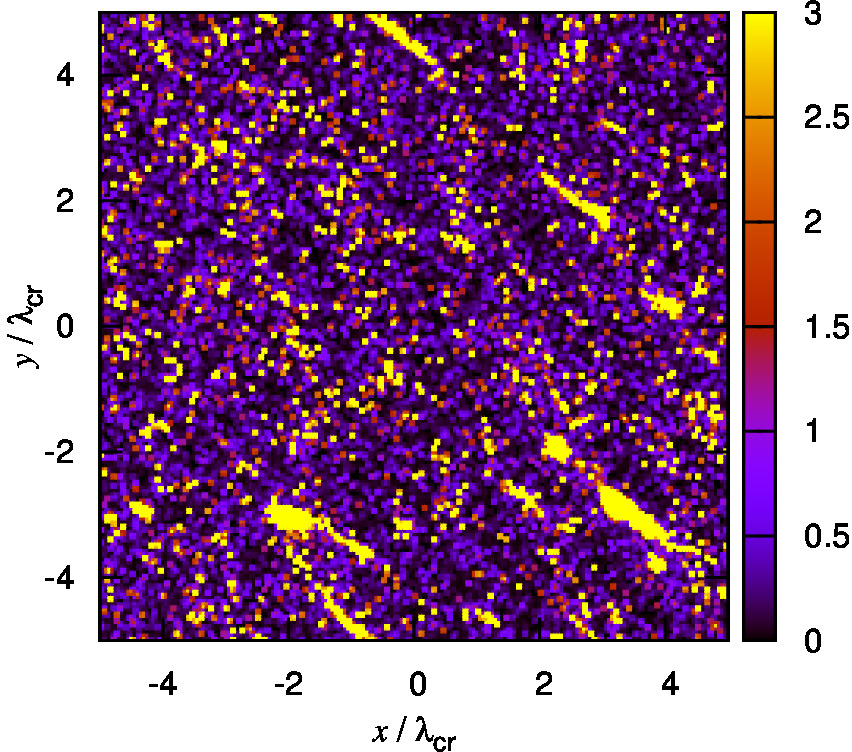}{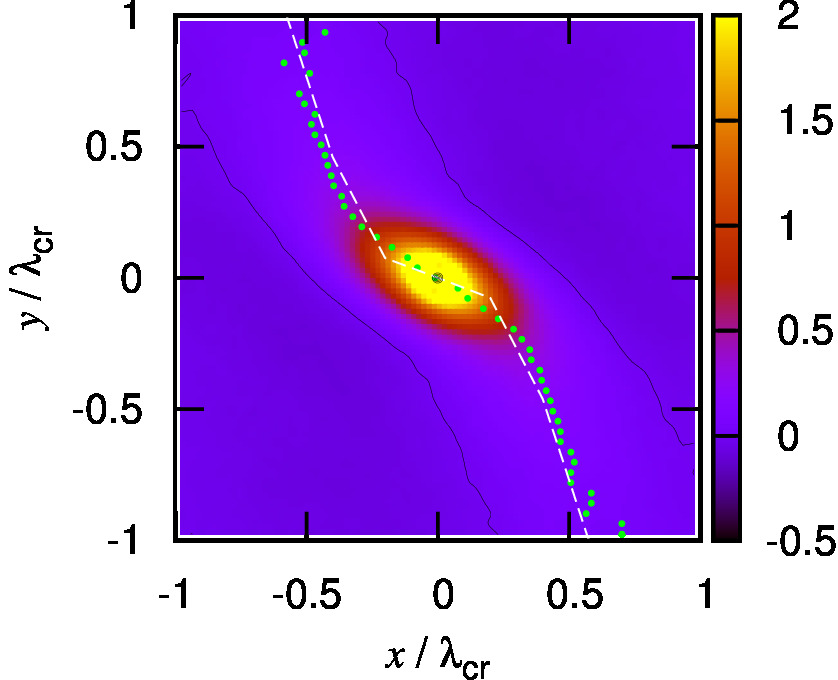}
 \caption{
   Surface density snapshot at $\tilde t=20.0$ (left) and its time-averaged autocorrelation (right)
   with the extracted ridge (dotted) and the fitting function (dashed)  for model 84. 
}
\label{fig:snapa135}
\end{figure}

For $\rho_\mathrm{p}=0.9 \, \mathrm{g} \, \mathrm{cm}^{-3}$ and $a \geq 1.3 \times 10^5 \, \mathrm{km}$, the inner pitch angle is about $90^\circ$.
Figure \ref{fig:snapa135} presents the snapshot of the surface density and the autocorrelation for $a = 1.35 \times 10^5 \, \mathrm{km}$ and $\rho_\mathrm{p} = 0.9 \, \mathrm{g} \, \mathrm{cm}^{-3}$ (model 84). 
In the surface density snapshot, we cannot observe the distinct elongated wakes but nearly spherical aggregates.
This is because the aggregate formation starts from $a\simeq 1.3 \times 10^5 \, \mathrm{km}$ for $\rho_\mathrm{p} = 0.9 \, \mathrm{g} \, \mathrm{cm}^{-3}$ \citep{Ohtsuki1993a, Salo1995, Karjalainen2004}.
In the autocorrelation function, the shape of the dense part is an ellipsoid, which corresponds to the Hill sphere. 
The long axis of the ellipsoid is almost parallel to the $x$-axis.
Thus, in this case, the inner pitch angle becomes around $90^\circ$.
The typical mass of the self-gravity wake is $\sim \Sigma_0 \lambda_\mathrm{cr}^2$. Thus, supposing that a particle with mass $\Sigma_0 \lambda_\mathrm{cr}^2$ is located at the center, we obtain the corresponding Hill radius $(\Sigma_0 \lambda_\mathrm{cr}^2 / 3M_\mathrm{S})^{1/3} a = \lambda_\mathrm{cr} / (12 \pi^2)^{1/3} \simeq 0.2 \lambda_\mathrm{cr}$. 
  This length scale is consistent with the size of the ellipsoid in autocorrelation function.

We compare the simulation results with the observations.
\cite{Hedman2007} found that the pitch angle increases from $17^\circ$ to $26^\circ$ with the Saturnicentric distance in the A ring.
\cite{Colwell2007} found the similar results that the pitch angle in the
A ring is $10 \mbox{--} 50 ^\circ$ and increases with the Saturnicentric distance. 
Our simulations with $\rho_\mathrm{p}=0.45\, \mathrm{g}\, \mathrm{cm}^{-3}$ show that the inner pitch angle increases 
with the Saturnicentric distance around $a \simeq 1.3 \times 10^5\, \mathrm{km}$ and 
the inner and outer pitch angles are  $43.0^\circ$ and $14.6^\circ$.
We expect that in general the observed pitch angle ranges within the inner and outer pitch angles.
Therefore, our simulations with $\rho_\mathrm{p}=0.45\, \mathrm{g}\, \mathrm{cm}^{-3}$ are consistent with the observations.

\section{Discussion \label{sec:dis}}

\subsection{Comparison with Galactic Spirals \label{sec:disgal}}

We compare the pitch angles of spiral arms in disk galaxies and
 self-gravity wakes in planetary rings.
Disk galaxies are considered as collisionless systems. 
Recently, the pitch angle of spiral arms was investigated by the numerical and analytic studies \citep{Grand2013, Michikoshi2014}. 
In the galactic disk, the pitch angle depends only on the shear rate defined by 
\begin{equation}
  	\Gamma = 2 - \frac{\kappa^2}{2 \Omega^2},
	\label{eq:shearrate}
\end{equation}
 where $\kappa$ is the epicycle frequency.
 From the numerical simulations and the linear analysis \cite{Michikoshi2014} obtained the fitting
 formula of the pitch angle
\begin{equation}
  \tan \theta = \frac{2}{7} \frac{\sqrt {4 - 2\Gamma}}{\Gamma}.
  \label{eq:pitch_collisionless}
\end{equation}
For the shear rate of the Keplerian rotation $\Gamma=3/2$, we obtain the pitch angle $\theta \simeq 11 ^\circ$,
which is consistent with the outer pitch angle of self-gravity wakes.

In the central dense parts where self-gravity wakes are closely packed, inelastic collisions play an important role in dynamics.
Thus, the inner pitch angle of self-gravity wakes deviates from the estimated value for collisionless systems.
In the outer low density regions, however, the effect of the shear is relatively strong compared to that of collisions, and the shear rate mainly determines the outer pitch angle just like collisionless systems.
Thus, the outer pitch angle is the same as that of collisionless systems and is independent of any ring parameters.

\subsection{Inner Pitch Angle \label{sec:dest}}
We simplify the physical processes concerning the inner pitch angle and derive a rough estimate as follows.
In high density regions, the self-gravity is strong.
Without the shear and only with the self-gravity, an ellipsoidal aggregate whose shape is the Hill sphere would form and its long axis is parallel to $x$-axis, that is, the pitch angle is $90^\circ$.
Therefore, the strong self-gravity tends to increase the inner pitch angle to $90^\circ$.
Conversely, the shear rotates the aggregate and tends to decrease the inner pitch angle to $0^\circ$.
The inner pitch angle may be determined by the balance between these two processes.

We consider the evolution of an aggregate formed due to the self-gravity.
When the aggregate is deformed by the shear, the self-gravity tends to restore it back to its original shape.
The restoration timescale may be approximated by $t_\mathrm{grav}\sim 1/\sqrt{G \rho_\mathrm{w}}$ where $\rho_\mathrm{w}$ is the bulk density of self-gravity wakes, 
which is the timescale of the Jeans instability or gravitational instability \citep[e.g.,][]{Chandrasekhar1953}.
The bulk density $\rho_\mathrm{w}$ is less than the particle density $\rho_\mathrm{p}$.
Introducing a dimensionless factor $C_\mathrm{w}$, we assume $\rho_\mathrm{w} = C_\mathrm{w} \rho_\mathrm{p}$. 
For identical particles, $C_\mathrm{w}$ is less than $0.74$ that is for the closest packing.
The timescale of self-gravity is given by
\begin{equation}
 t_\mathrm{grav}  \simeq 
 \frac{ C_\mathrm{d} }{\sqrt{G C_\mathrm{w} \rho_\mathrm{p}}} = 
 \frac{C_\mathrm{d}}{3 \Omega } \sqrt{\frac{\pi}{C_\mathrm{w}}}
 \tilde r_\mathrm{H}^{-3/2},
 \label{eq:stime}
\end{equation}
 where $C_\mathrm{d}$ is a non-dimensional factor on the order of unity.
The shear rotates and extends the wake and the pitch angle decreases with $t$ as
\begin{equation}
  \tan \theta = \frac{2}{3 \Omega t},
\label{eq:shear}
\end{equation}
where $t$ is the time elapsed since $\theta=90^\circ$ \citep[e.g.,][]{Toomre1981, Binney2008}.  The shear timescale is
\begin{equation} 
 t_\mathrm{shear} \simeq
 \left|\frac{1}{\tan \theta}\frac{\mathrm{d}\tan \theta}{\mathrm{d}t}
 \right|^{-1} = \frac {2}{3 \Omega \tan \theta}. 
 \label{eq:pitch}
\end{equation} 
As the pitch angle decreases with time, the wake rotation slows down.

When the wake pitch angle is large ($\theta \sim 90^\circ$), the
 wake rotation is fast ($t_\mathrm{shear} < t_\mathrm{grav}$). 
The wake rotates quickly before the self-gravity restores the inclined wake back to its original shape.
As the pitch angle decreases, the wake rotation slows down and the self-gravity becomes important. 
Thus, from $t_\mathrm{shear} = t_\mathrm{grav}$ we obtain a crude estimate of the inner pitch angle:
\begin{equation} 
  \tan \theta_\mathrm{i} \simeq 
 \frac{2}{C_\mathrm{d}} \sqrt{\frac{C_\mathrm{w}}{\pi}} 
 \tilde r_\mathrm{H}^{3/2}
 = 0.84 \frac{\sqrt{C_\mathrm{w}}}{C_\mathrm{d}} 
 \left(\frac{\rho_\mathrm{p}}{0.9\, \mathrm{g}\, \mathrm{cm}^{-3}} \right)^{1/2}
 \left(\frac{a}{10^{5}\, \mathrm{km}} \right)^{3/2}
 , 
 \label{eq:theory}
\end{equation} 
which means that the pitch angle increases with $C_\mathrm{w}$. 
The bulk density of self-gravity wakes with size distribution can be larger than that for identical particles, and thus the pitch angle with size distribution is larger than that for identical particles, which is consistent with the simulation results. 
Figure \ref{fig:pitch_rdep_size} presents the pitch angle as a function of $R_\mathrm{max}/R_\mathrm{min}$.
As $R_\mathrm{max}/R_\mathrm{min}$ increases, the filling factor $C_\mathrm{w}$ increases. 
Assuming $C_\mathrm{d}=1.0$, 
we find that $C_\mathrm{w}=0.45$ and $1$ are suitable for the identical particle models and the size distribution models of $R_\mathrm{max}/R_\mathrm{min}=10$, respectively. 
As shown in Figures \ref{fig:pitch_rdep} and \ref{fig:pitch_sizea}, Equation (\ref{eq:theory}) can roughly explain the dependence of
the inner pitch angle on $\tilde r_\mathrm{H}$ and $a$ with $C_\mathrm{w} = 0.45$ for the identical particle models and $C_\mathrm{w} = 1.0$ for the size distribution models with $R_\mathrm{max}/R_\mathrm{min}=10$.
The simulation results and Equation (\ref{eq:theory}) show that the inner pitch angle increases with $\tilde r_\mathrm{H}$.  
Since $\tilde r_\mathrm{H}$ is proportional to the Saturnicentric distance
 $a$, as $\tilde r_\mathrm{H}$ increases, the orbital period increases, that is, the shear timescale increases. 
On the other hand, the timescale of the gravitational instability depends only on $\rho_\mathrm{w}$ and is independent of $a$.  
Therefore, as $\tilde r_\mathrm{H}$ increases, the self-gravity dominates over the shear and the inner pitch angle increases.

Note that the actual filling factor has not been investigated in detail yet. 
  These results do not suggest that the filling factor is $C_\mathrm{w}=0.45$ or $1.0$.  The factor $C_\mathrm{d}$ is also
  a free parameter. If we adopt smaller $C_\mathrm{d}$, the
  corresponding filling factor becomes smaller.  The realistic filling factor
  should be smaller than unity.

We used $R_\mathrm{max}/ R_\mathrm{min} \le 10$ in the size distribution models. 
However, this size ratio is far smaller than the observationally inferred values that are $70$ and $2000$ in the inner and outer A rings, respectively \citep{French2000}.
As shown in Figure \ref{fig:pitch_rdep_size}, the inner pitch angle is almost constant for $R_\mathrm{max}/ R_\mathrm{min}>5$.
Accordingly, the simulation results with $R_\mathrm{max}/ R_\mathrm{min}=10$ in Section \ref{secsize} may be applicable to the realistic ring.

\section{Summary}

We have investigated the pitch angle of the self-gravity wakes in the dense planetary ring by performing local $N$-body simulations where both inelastic collisions and self-gravitational interactions between particles are included. 
In order to extract the feature of the self-gravity wakes, we calculated the two-dimensional autocorrelation function of the surface density.
We measured the angle of the inclined structure of the autocorrelation function as the wake pitch angle. 
We obtained two pitch angles in the inner and outer parts of the autocorrelation function separately.
We confirmed that the inner pitch angle is larger than the outer one in all models.

We examined the dependence of the pitch angle on various ring parameters.
We found that the outer pitch angle is independent of the ring parameters, which is $10^\circ\mbox{--}15^\circ$.
This value is consistent with the estimation of pitch angle of the spiral arms in collisionless systems \citep{Julian1966, Michikoshi2014}.
The inner pitch angle does not depend on the restitution coefficient $\epsilon$ of particles and the optical depth $\tau$ 
provided that the overstable oscillation does not occur.
We found that the inner pitch angle is determined by the Hill radius of a particle pair relative to the sum of their physical radii $\tilde r_\mathrm{H}$.
The inner pitch angle increases with $\tilde r_\mathrm{H}$ from $20^\circ$ to $50^\circ$.
The tendency that the pitch angle increases with $\tilde r_\mathrm{H}$ is consistent with the result of the observational studies that the pitch angle increases with the Saturnicentric distance \citep{Hedman2007, Colwell2007}.  
The pitch angle with the size distribution model is larger than that for the identical particles because the filling factor for the size distribution model is larger and the resulting self-gravity is stronger.
The strong self-gravity increases the inner pitch angle.

Considering the effects of the self-gravity and the shear, the tendency of the increase of the inner pitch angle with the Saturnicentric distance can be explained.
However, this model is far from a complete theory.
We need a more sophisticated theoretical framework to understand the overall activity of the self-gravity wakes and more detailed analyses of the self-gravity wakes by $N$-body simulation.
It is especially important to understand the wake amplification and destruction processes considering the inelastic collisions and the nonlinear effects.
We will focus on these processes in the next paper.

\acknowledgments{
Numerical computations were carried out on the GRAPE system at Center for Computational Astrophysics, National Astronomical Observatory of Japan.
}

\end{document}